\documentclass[english]{IEEEtran}
\usepackage[T1]{fontenc}
\usepackage[latin9]{inputenc}
\usepackage{float}
\usepackage{amsthm}
\usepackage{amsmath}
\usepackage{amssymb}
\usepackage{graphicx}

\makeatletter

\floatstyle{ruled}
\newfloat{algorithm}{tbp}{loa}
\providecommand{\algorithmname}{Algorithm}
\floatname{algorithm}{\protect\algorithmname}

\theoremstyle{plain}
\newtheorem{thm}{\protect\theoremname}
\theoremstyle{plain}
\newtheorem{lem}[thm]{\protect\lemmaname}
\theoremstyle{definition}
\newtheorem{defn}[thm]{\protect\definitionname}
\theoremstyle{remark}
\newtheorem{rem}[thm]{\protect\remarkname}
\theoremstyle{plain}
\newtheorem{cor}[thm]{\protect\corollaryname}
\theoremstyle{definition}
\newtheorem{example}[thm]{\protect\examplename}
\theoremstyle{plain}
\newtheorem{prop}[thm]{\protect\propositionname}

\usepackage{cite}

\@ifundefined{showcaptionsetup}{}{%
 \PassOptionsToPackage{caption=false}{subfig}}
\usepackage{subfig}
\makeatother

\usepackage{babel}
\providecommand{\corollaryname}{Corollary}
\providecommand{\definitionname}{Definition}
\providecommand{\examplename}{Example}
\providecommand{\lemmaname}{Lemma}
\providecommand{\propositionname}{Proposition}
\providecommand{\remarkname}{Remark}
\providecommand{\theoremname}{Theorem}

\begin{document}

\title{Regularized Tyler's Scatter Estimator: Existence, Uniqueness, and
Algorithms}

\author{Ying~Sun,~Prabhu~Babu,~and~Daniel~P.~Palomar,~\IEEEmembership{Fellow,~IEEE}%
\thanks{Ying~Sun,~Prabhu~Babu,~and~Daniel~P.~Palomar are with the Hong
Kong University of Science and Technology (HKUST), Hong Kong. E-mail:
\{ysunac,~eeprabhubabu,~palomar\}@ust.hk.%
}%
\thanks{This work was supported by the Hong Kong RGC 617312 research grant.%
}}
\maketitle
\begin{abstract}
This paper considers the regularized Tyler's scatter estimator for
elliptical distributions, which has received considerable attention
recently. Various types of shrinkage Tyler's estimators have been
proposed in the literature and proved work effectively in the ``small
n large p'' scenario. Nevertheless, the existence and uniqueness
properties of the estimators are not thoroughly studied, and in certain
cases the algorithms may fail to converge. In this work, we provide
a general result that analyzes the sufficient condition for the existence
of a family of shrinkage Tyler's estimators, which quantitatively
shows that regularization indeed reduces the number of required samples
for estimation and the convergence of the algorithms for the estimators.
For two specific shrinkage Tyler's estimators, we also proved that
the condition is necessary and the estimator is unique. Finally, we
show that the two estimators are actually equivalent. Numerical algorithms
are also derived based on the majorization-minimization framework,
under which the convergence is analyzed systematically.\end{abstract}

\section{Introduction}

Covariance estimation has been a long existing problem in various
signal processing related fields, including multiantenna communication
systems, social networks, bioinformatics, as well as financial engineering.
A well known and easy to implement estimator is the sample covariance
matrix. Under the assumption of clean samples, the estimator is consistent
by the Law of Large Numbers. However, the performance of the sample
covariance matrix is vulnerable to data corrupted by noise and outliers,
which is often the case in real-world applications.

As a remedy, robust estimators are proposed aimed at limiting the
influence of erroneous observations so as to achieve better performance
in non-Gaussian scenarios \cite{maronna2006robust,huber2004robust}.
Recently, Tyler's scatter estimator \cite{tyler1987distribution}
has received considerable attention both theoretically and practically
in signal processing related fields, e.g., \cite{Ollila2008,Ollila2009,Ollila2012a,Pascal2008,wiesel2012geodesic}
to name a few, see \cite{Ollila2012} for a comprehensive overview.
Tyler's estimator estimates the normalized scatter matrix (equivalently
the normalized covariance matrix if the covariance exists) assuming
that the underlying distribution is elliptically symmetric. The estimator
is shown to enjoy the following advantages against the others: it
is distribution-free in the sense that its asymptotic variance does
not depend on the parametric form of the underlying distribution,
and it is also the most robust estimator in a min-max sense.

In addition to non-Gaussian observations, another problem we face
in practice is the ``small n large p'' problem, which refers to
high dimensional statistical inference with insufficient number of
samples. It is obvious that the sample covariance matrix is singular
when the number of samples is smaller than the dimension, and Tyler's
estimator has the same drawback. In order to handle this problem,
\cite{4217907} borrowed the diagonal loading idea \cite{ledoit2004well}
and proposed a regularized Tyler's estimator that shrinks towards
identity. A rigorous proof for the existence, uniqueness, and convergence
properties is provided in \cite{chen2011robust}, where a systematic
way of choosing the regularization parameter was also proposed. However,
the estimator is criticized for not being derived from a meaningful
cost function. To overcome this issue, a new scale-invariant shrinkage
Tyler's estimator, defined as a minimizer of a penalized cost function,
was recently proposed in \cite{wiesel2012unified}. By showing that
the objective function is geodesic convex, Wiesel proved that any
algorithm that converges to the local minimum of the objective function
is actually the global minimum. Numerical algorithms are provided
for the estimator and simulation results demonstrate the estimator
is robust and effective in the sample deficient scenario. Despite
the good properties, the existence and uniqueness properties of the
estimator remains unclear.

In this paper, we study the shrinkage Tyler's estimator and try to
answer the unsolved problems mentioned above. First, we give a proof
that states the sufficient condition for the existence of shrinkage
Tyler's estimator with penalized cost function taking a general form.
Second, we propose a Kullback-Leibler divergence (KL divergence) penalized
cost function that results in a shrinkage Tyler's estimator similar
to the heuristic diagonal loading one considered in \cite{4217907,chen2011robust}.
We then move to these two specific estimators and show that under
the condition $P_{N}\left(S\right)<\frac{\left(1+\alpha_{0}\right){\rm dim}\left(S\right)}{K}$
and the shrinkage target matrix being positive definite, the estimators
exist, where $N$ is the number of samples, $K$ is the dimension
of the samples and $\alpha_{0}$ controls the amount of penalty added
to the cost function, $P_{N}\left(S\right)$ stands for the proportion
of samples contained in a proper subspace $S$. In addition, we prove
it is also a necessary condition, provided that $\alpha_{0}>0$. Although
derived from different cost functions, and also with different estimation
equation, we prove that the two shrinkage estimators are actually
equivalent. Under the assumption that the underlying distribution
is continuous, the condition simplifies to $N>\frac{K}{1+\alpha_{0}}$.
Comparing with the existence condition for Tyler's estimator, which
is $P_{N}\left(S\right)<\frac{{\rm dim}\left(S\right)}{K}$, or $N>K$
under continuity assumption, this result clearly demonstrates that
regularization can relax the requirement on the number of samples,
hence shows its capability of handling large dimension estimation
problems. Algorithms for the shrinkage estimators based on majorization-minimization
framework are provided, where the convergence can be analyzed systematically.

It is worth mentioning that in the work \cite{pascal2013generalized},
where the same condition $N>\frac{K}{1+\alpha_{0}}$ is also independently
derived for the KL penalty based shrinkage estimator that shrinks
the covariance matrix to identity in the complex field, assuming the
samples are linearly independent. \cite{pascal2013generalized} refutes
the additional trace normalization step in \cite{chen2011robust}
by showing that the trace of the inverse of the estimator is equal
to $K$, and propose dropping the normalization step. Different from
that approach, our work gives an interpretation of the estimator as
the minimizer of a KL divergence penalized cost function. Starting
from the cost function, we establish the existence condition with
a different proof from \cite{pascal2013generalized}. In addition,
we extend the result (in the real field), since the condition $P_{N}\left(S\right)<\frac{\left(1+\alpha_{0}\right){\rm dim}\left(S\right)}{K}$
implies $N>\frac{K}{1+\alpha_{0}}$ if the samples are linearly independent,
and we consider a general positive definite shrinkage target matrix
as in \cite{wiesel2012unified}.

The paper is organized as follows: In Section II, we briefly review
Tyler's estimator for samples drawn from the elliptical family. In
Section III, the two types of shrinkage estimators, i.e., one proposed
in \cite{wiesel2012unified} and another derived based on KL divergence
are considered, and a rigorous proof for the existence and uniqueness
of the estimators is provided. Algorithms based on majorization-minimization
are presented in Section IV. Numerical examples follow in Section
V, and we conclude in Section VI.

\subsection*{Notation}

$\mathbb{R}^{n}$ stands for $n$-dimensional real-valued vector space,
$\left\Vert \cdot\right\Vert _{2}$ stands for vector Frobenius norm.
$\mathbb{S}_{+}^{K}$ stands for symmetric positive semidefinite $K\times K$
matrices, which is a closed cone in $\mathbb{R}^{K\times K}$, $\mathbb{S}_{++}^{K}$
denotes symmetric positive definite $K\times K$ matrices. $\lambda_{\max}$
and $\lambda_{\min}$ stand for the largest and smallest eigenvalue
of a matrix $\boldsymbol{\Sigma}$ respectively. $\det\left(\cdot\right)$
and ${\rm Tr}\left(\cdot\right)$ stand for matrix determinant and
trace respectively. $\left\Vert \cdot\right\Vert _{F}$ is the matrix
Frobenius norm.

The boundary of the open set $\mathbb{S}_{++}^{K}$ is conventionally
defined as $\mathbb{S}_{+}^{K}\backslash\mathbb{S}_{++}^{K}$, which
contains all rank deficient matrices in $\mathbb{S}_{+}^{K}$. With
a slightly abuse of notation, we also include matrices with all eigenvalues
$\lambda\to+\infty$ into the boundary of $\mathbb{S}_{++}^{K}$.
Therefore a sequence of matrices $\boldsymbol{\Sigma}^{k}$ converges
to the boundary of $\mathbb{S}_{++}^{K}$ iff $\lambda_{\max}^{k}\to+\infty$
or $\lambda_{\min}^{k}\to0$. In the rest of the paper, we will use
the statement ``$\boldsymbol{\Sigma}$ converges'' equivalently
as ``a sequence of matrices $\boldsymbol{\Sigma}^{k}$ converges''
for notation simplicity.

\section{Robust covariance matrix estimation}

In this paper, we assume a number $N$ of $K$-dimensional samples
$\left\{ {\bf x}_{1},\ldots,{\bf x}_{N}\right\} $ are drawn from
an elliptical population distribution with probability density function
(pdf) of the form
\begin{equation}
f\left({\bf x}\right)=\det\left(\boldsymbol{\Sigma}_{0}\right)^{-\frac{1}{2}}g\left(\left({\bf x}-\boldsymbol{\mu}_{0}\right)^{T}\boldsymbol{\Sigma}_{0}^{-1}\left({\bf x}-\boldsymbol{\mu}_{0}\right)\right)\label{eq:elliptical pdf}
\end{equation}
with location and scatter parameter $\left(\boldsymbol{\mu}_{0},\boldsymbol{\Sigma}_{0}\right)$
in $\mathbb{R}^{K}\times\mathbb{S}_{++}^{K}$. The nonnegative function
$g\left(\cdot\right)$, which is called the density generator, determines
the shape of the pdf. In most of the popularly used distributions,
e.g., the Gaussian and the Student's \textsl{t}-distribution, $g\left(\cdot\right)$
is a decreasing function and determines the decay of the tails of
the distribution. Given $\boldsymbol{\mu}_{0}$, our problem of interest
is to estimate the covariance matrix. We can always center the pdf
by defining $\tilde{{\bf x}}={\bf x}-\boldsymbol{\mu}_{0}$, hence
without loss of generality in the rest of the paper we assume $\boldsymbol{\mu}_{0}={\bf 0}.$
We use the notation $P_{N}$ and $f\left({\bf \cdot}\right)$ for
the empirical and the population distributions, respectively. It is
known that the covariance matrix of elliptical distribution takes
the form $c_{g}\boldsymbol{\Sigma}_{0}$ with $c_{g}$ being a constant
that depends on $g\left(\cdot\right)$ \cite{maronna2006robust},
hence it is unlikely to have a good covariance estimator without prior
knowledge of $g$. In this paper, instead of trying to find the parametric
form of $g$ and get an estimator of $c_{g}\boldsymbol{\Sigma}_{0}$,
we are interested in estimating the normalized covariance matrix $\frac{\boldsymbol{\Sigma}_{0}}{{\rm Tr}\left(\boldsymbol{\Sigma}_{0}\right)}$.

The commonly used sample covariance matrix, which also happens to
be the maximum likelihood estimator for the normal distribution, estimates
$c_{g}\boldsymbol{\Sigma}_{0}$ asymptotically, however it is sensitive
to outliers. This motivates the research for estimators robust to
outliers in the data and, in fact, many researchers in the statistics
literature have addressed this problem by proposing various robust
covariance estimators like M-estimators \cite{maronna1976robust},
S-estimators \cite{davies1987asymptotic}, MVE \cite{van2009minimum},
and MCD \cite{butler1993asymptotics} to name a few, see \cite{maronna2006robust,huber2004robust}
for a complete overview. For example, in \cite{maronna1976robust},
Maronna analyzed the properties of the M-estimators, which are given
as the solution $\boldsymbol{\Sigma}$ to the equation
\begin{equation}
\boldsymbol{\Sigma}=\frac{1}{N}\sum_{i=1}^{N}u\left({\bf x}_{i}^{T}\boldsymbol{\Sigma}^{-1}{\bf x}_{i}\right){\bf x}_{i}{\bf x}_{i}^{T}\label{eq:Maronna Scatter}
\end{equation}
where the choice of function $u\left(\cdot\right)$ determines a whole
family of different estimators. Under some technical conditions on
$u\left(s\right)$ (i.e., $u\left(s\right)\geq0$ for $s>0$ and nonincreasing,
and $su\left(s\right)$ is strictly increasing), Maronna proved that
there exists a unique $\boldsymbol{\Sigma}$ that solves (\ref{eq:Maronna Scatter}),
and gave an iterative algorithm to arrive at that solution. He also
established its consistency and robustness. A number of well known
estimators take the form (\ref{eq:Maronna Scatter}) and in \cite{maronna1976robust}
Maronna gave two examples, with one being the maximum likelihood estimator
for multivariate Student's \textsl{t}-distribution, and the other
being the Huber's estimator \cite{huber1964robust}. Both of them
are popular for handling heavy tails and outliers in the data.

For all the robust covariance estimators, there is a tradeoff between
their efficiency, which measures the variance (estimation accuracy)
of the estimator, and robustness, which quantifies the sensitivity
of the estimator to outliers. As these two quantities are opposed
in nature, a considerable effort has to be put in designing estimators
that achieve the right balance between these two quantities. In \cite{tyler1987distribution},
Tyler dealt with this problem by proposing an estimator that is distribution-free
and the ``most robust'' estimator in mini-max sense. Tyler's estimator
of $\boldsymbol{\Sigma}$ is given as the solution of the following
equation
\begin{equation}
\boldsymbol{\Sigma}=\frac{K}{N}\sum_{i=1}^{N}\frac{{\bf x}_{i}{\bf x}_{i}^{T}}{{\bf x}_{i}^{T}\boldsymbol{\Sigma}^{-1}{\bf x}_{i}}\label{eq:Tyler Scatter}
\end{equation}
where the results of \cite{maronna1976robust} cannot be applied since
$su\left(s\right)=K$ is not strictly increasing. Tyler established
the conditions for the existence of a solution to the fixed-point
equation (\ref{eq:Tyler Scatter}), as well as the fact that the estimator
is unique up to a positive scaling factor, in the sense that $\boldsymbol{\Sigma}$
solves (\ref{eq:Tyler Scatter}) if and only if $c\boldsymbol{\Sigma}$
solves (\ref{eq:Tyler Scatter}) for some positive scalar $c$. The
estimator was shown to be strongly consistent and asymptotically normal
with its asymptotic standard deviation independent of $g$.

Tyler's fixed-point equation (\ref{eq:Tyler Scatter}) can be alternatively
interpreted as follows. Consider the normalized samples defined as
${\bf s}=\frac{{\bf x}}{\left\Vert {\bf x}\right\Vert _{2}}$, it
is known that the probability distribution of ${\bf s}$ takes the
form \cite{tyler1987statistical,kent1988maximum,frahm2004generalized}
\begin{equation}
f\left({\bf s}\right)=\frac{\Gamma\left(\frac{K}{2}\right)}{2\pi^{K/2}}\det\left(\boldsymbol{\Sigma}\right)^{-\frac{1}{2}}\left({\bf s}^{T}\boldsymbol{\Sigma}^{-1}{\bf s}\right)^{-K/2}.\label{eq:pdf-normalized-elliptic}
\end{equation}
Given $N$ samples from the normalized distribution $\left\{ {\bf s}_{i}\right\} $,
the maximum likelihood estimator of $\boldsymbol{\Sigma}$ can be
obtained by minimizing the negative log-likelihood function
\begin{equation}
L\left(\boldsymbol{\Sigma}\right)=\sum_{i=1}^{N}\frac{K}{2}\log\left({\bf s}_{i}^{T}\boldsymbol{\Sigma}^{-1}{\bf s}_{i}\right)+\frac{N}{2}\log\det\left(\boldsymbol{\Sigma}\right)\label{eq:}
\end{equation}
which is equivalent to minimizing
\begin{equation}
L^{{\rm Tyler}}\left(\boldsymbol{\Sigma}\right)=\sum_{i=1}^{N}\frac{K}{2}\log\left({\bf x}_{i}^{T}\boldsymbol{\Sigma}^{-1}{\bf x}_{i}\right)+\frac{N}{2}\log\det\left(\boldsymbol{\Sigma}\right).\label{eq:Loss function Sigma}
\end{equation}
If a minimum $\hat{\boldsymbol{\Sigma}}\succ{\bf 0}$ of the function
$L^{{\rm Tyler}}\left(\boldsymbol{\Sigma}\right)$ exists, it needs
to satisfy the stationary equation given in (\ref{eq:Tyler Scatter}),
which was originally derived by Tyler in \cite{tyler1987distribution}.
In \cite{tyler1987distribution,kent1988maximum}, the authors provided
the condition for existence of a nonsingular solution to (\ref{eq:Tyler Scatter})
based on the following reasoning. Notice that $\hat{\boldsymbol{\Sigma}}$
must be nonsingular, and the function $L^{{\rm Tyler}}\left(\boldsymbol{\Sigma}\right)$
is unbounded above on the boundary of positive definite matrices,
implies the existence of a minimum. Based on these observations, Kent
and Tyler established the existence conditions by showing $L^{{\rm Tyler}}\left(\boldsymbol{\Sigma}\right)\to+\infty$
on the boundary. Specifically, under the condition that: (i) no ${\bf x}_{i}$
lies on the origin, and (ii) for any proper subspace $S\subseteq\mathbb{R}^{K}$,
$P_{N}\left(S\right)<\frac{{\rm dim}\left(S\right)}{K}$, where $P_{N}\left(S\right)\triangleq\frac{\sum_{i=1}^{N}1_{\left\{ {\bf x}_{i}\in S\right\} }}{N}$
stands for the proportion of samples in $S$, then a nonsingular minimum
of the problem (\ref{eq:Loss function Sigma}) exists, which is equivalent
to equation (\ref{eq:Tyler Scatter}) having a solution. In words,
the above mentioned conditions require the number of samples to be
sufficiently large, and the samples should be spread out in the whole
space.

To arrive at the estimator satisfying (\ref{eq:Tyler Scatter}), Tyler
proposed the following iterative algorithm:
\begin{equation}
\begin{array}{l}
\tilde{\boldsymbol{\Sigma}}_{t+1}=\frac{K}{N}\sum_{i=1}^{N}\frac{{\bf x}_{i}{\bf x}_{i}^{T}}{{\bf x}_{i}^{T}\boldsymbol{\Sigma}_{t}^{-1}{\bf x}_{i}}\\
\boldsymbol{\Sigma}_{t+1}=\frac{\tilde{\boldsymbol{\Sigma}}_{t+1}}{{\rm Tr}\left(\tilde{\boldsymbol{\Sigma}}_{t+1}\right)}
\end{array}\label{eq:Alg-Tyler Scatter}
\end{equation}
that converges to the unique (up to a positive scaling factor) solution
of (\ref{eq:Tyler Scatter}).

The robust property of Tyler's estimator can be understood intuitively
as follows: by normalizing the samples, i.e., ${\bf s}=\frac{{\bf x}}{\left\Vert {\bf x}\right\Vert _{2}}$,
the magnitude of an outlier is more unlikely to make the estimator
break down. In other words, the estimator is not sensitive to the
magnitude of samples, only their direction can affect the performance.

\section{Regularized Covariance Matrix Estimation}

The regularity conditions for the existence of Tyler's estimator leads
to a condition on the number of samples that $N\geq K+1$ \cite{kent1988maximum,kent1991redescending}.
In some practical applications the number of samples is not sufficient,
in those cases Tyler's iteration (\ref{eq:Alg-Tyler Scatter}) may
not converge. In these scenarios, a most sensible approach is to shrink
Tyler's estimator to some known a priori estimate of $\boldsymbol{\Sigma}$.
In the literature of robust estimators, there exists two different
shrinkage based approaches.

In the first approach, the authors in \cite{4217907,chen2011robust}
proposed the following estimator:
\begin{equation}
\begin{array}{l}
\tilde{\boldsymbol{\Sigma}}_{t+1}=\frac{1}{1+\alpha_{0}}\frac{K}{N}\sum_{i=1}^{N}\frac{{\bf x}_{i}{\bf x}_{i}^{T}}{{\bf x}_{i}^{T}\boldsymbol{\Sigma}_{t}^{-1}{\bf x}_{i}}+\frac{\alpha_{0}}{1+\alpha_{0}}{\bf I}\\
\boldsymbol{\Sigma}_{t+1}=\frac{\tilde{\boldsymbol{\Sigma}}_{t+1}}{{\rm Tr}\left(\tilde{\boldsymbol{\Sigma}}_{t+1}\right)}
\end{array}\label{eq:heuristic diagonal loading}
\end{equation}
which is a slightly modified version of the original Tyler's iteration
in (\ref{eq:Alg-Tyler Scatter}), with the modification being including
an identity matrix in the first step of the iteration that aims at
shrinking the estimator towards the identity matrix. This resembles
the idea of regularizing an estimator via diagonal loading \cite{ledoit2004well,6576912}.
In \cite{chen2011robust}, Chen et al. proved the uniqueness of the
estimator obtained by the iteration (\ref{eq:heuristic diagonal loading})
based on concave Perron-Frobenius theory, and gave a method to choose
the regularization weight $\alpha_{0}$. Although this estimator is
widely used and performs well in practice, it is still considered
to be heuristic as it does not have an interpretation based on minimizing
a cost function.

As a second approach, in \cite{wiesel2012unified}, the author took
a different route and derived a new shrinkage-based Tyler's estimator
that has a clear interpretation based on minimizing the penalized
negative log-likelihood function

\begin{equation}
L^{{\rm Wiesel}}\left(\boldsymbol{\Sigma}\right)=\frac{2}{N}L^{{\rm Tyler}}\left(\boldsymbol{\Sigma}\right)+\alpha_{0}h^{{\rm target}}\left(\boldsymbol{\Sigma}\right)\label{eq:Wiesel penalty general objective}
\end{equation}
where $h^{{\rm target}}\left(\boldsymbol{\Sigma}\right)=K\log\left({\rm Tr}\left(\boldsymbol{\Sigma}^{-1}{\bf T}\right)\right)+\log\det\left(\boldsymbol{\Sigma}\right)$
is a function with minimum at the desired target matrix ${\bf T}$,
hence it will shrink the solution of (\ref{eq:Wiesel penalty general objective})
towards the target. By showing the cost function $L^{{\rm Wiesel}}\left(\boldsymbol{\Sigma}\right)$
is geodesic convex, the author proved that any local minimum over
the set of positive definite matrices is a global minimum \cite{wiesel2012unified}.
He then derived an iterative algorithm based on majorization-minimization
that monotonically decreases the cost function at each iteration:
\begin{equation}
\begin{aligned}\boldsymbol{\Sigma}_{t+1} & =\frac{1}{1+\alpha_{0}}\frac{K}{N}\sum_{i=1}^{N}\frac{{\bf x}_{i}{\bf x}_{i}^{T}}{{\bf x}_{i}^{T}\boldsymbol{\Sigma}_{t}^{-1}{\bf x}_{i}}+\frac{\alpha_{0}}{1+\alpha_{0}}\frac{K{\bf T}}{{\rm Tr}\left(\boldsymbol{\Sigma}_{t}^{-1}{\bf T}\right)}.\end{aligned}
\label{eq:wiesel penality}
\end{equation}

Even though the author in \cite{wiesel2012unified} showed that the
cost function is convex in geodesic space, the existence and uniqueness
of the global minimizer remains unknown. Moreover, it is mentioned
in \cite{wiesel2012unified} that for some values of $\alpha_{0}$
the cost function becomes unbounded below and the iterations do not
converge.

In this section, we address the following points: (i) we give the
missing interpretation based on minimizing a cost function for the
estimator in (\ref{eq:heuristic diagonal loading}), and we also prove
its existence and uniqueness; (ii) we prove the iteration in (\ref{eq:wiesel penality})
with an additional trace normalization step converges to a unique
point and also establish the conditions on the regularization parameter
$\alpha_{0}$ to ensure the existence of the solution. For both cases,
the cost function takes the form of penalized negative log-likelihood
function with different penalizing functions. Our methodology for
the proofs hinges on techniques used by Tyler in \cite{kent1991redescending,kent1988maximum}.

We start with a proof of existence for a minimizer of a general penalized
negative log-likelihood function in the following theorem, the proof
of existence of the two aforementioned cases $L^{{\rm Tyler}}\left(\boldsymbol{\Sigma}\right)$
and $L^{{\rm Wiesel}}\left(\boldsymbol{\Sigma}\right)$ are just special
cases of the general result.

The idea of proving the existence is to establish the regularity conditions
under which the cost function takes value $+\infty$ on the boundary
of the set $\mathbb{S}_{++}^{K}$, a minimum then exists by the continuity
of the cost function. The main result is established in Theorem \ref{thm:Existence Thm},
and the following lemma is needed.
\begin{lem}
\label{lem:equivalence lem}For any continuous function $f\left(\cdot\right)$
defined on the set $\mathbb{S}_{++}$, there exists a $\hat{\boldsymbol{\Sigma}}\succ{\bf 0}$
such that $f\left(\hat{\boldsymbol{\Sigma}}\right)\leq f\left(\boldsymbol{\Sigma}\right)\ \forall\boldsymbol{\Sigma}\succ{\bf 0}$
if $f\left(\boldsymbol{\Sigma}\right)\to+\infty$ on the boundary
of the set $\mathbb{S}_{++}$.\end{lem}
\begin{defn}
\label{Def: als}For any continuous function $f\left(s\right)$ defined
on $s>0$, define the quantities
\begin{equation}
a_{f}=\sup\left\{ a|s^{a/2}\exp\left(-f\left(s\right)\right)\to0\ {\rm as}\ s\to+\infty\right\} \label{eq:sill}
\end{equation}
and
\begin{equation}
a_{f}^{'}=\inf\left\{ a|s^{a/2}\exp\left(-f\left(s\right)\right)\to0\ {\rm as}\ s\to0\right\} \label{eq:floor}
\end{equation}

\end{defn}
In this paper we are particularly interested in the functions $f\left(s\right)=c\log s$
and $f\left(s\right)=cs$ with some positive scalar $c<+\infty$.
For $f\left(s\right)=c\log s$, $a_{f}=a_{f}^{'}=2c$ and, for $f\left(s\right)=cs$,
$a_{f}=+\infty$, $a_{f}^{'}=0$. We restrict our attention to the
case $a_{f}\geq0$.

Consider the penalized cost function takes the general form $\tilde{L}\left(\boldsymbol{\Sigma}\right)=L^{\rho}\left(\boldsymbol{\Sigma}\right)+h\left(\boldsymbol{\Sigma}\right)$
with original cost function
\begin{equation}
\begin{array}{c}
L^{\rho}\left(\boldsymbol{\Sigma}\right)=\frac{N}{2}\log\det\left(\boldsymbol{\Sigma}\right)+\sum_{i=1}^{N}\rho\left({\bf x}_{i}^{T}\boldsymbol{\Sigma}^{-1}{\bf x}_{i}\right)\end{array}\label{eq:-1}
\end{equation}
where $\rho\left(\cdot\right)$ is a continuous function, and the
penalty term
\begin{equation}
h\left(\boldsymbol{\Sigma}\right)=\alpha\log\det\left(\boldsymbol{\Sigma}\right)+\sum_{l=1}^{L}\alpha_{l}h_{l}\left({\rm Tr}\left({\bf A}_{l}^{T}\boldsymbol{\Sigma}^{-1}{\bf A}_{l}\right)\right)\label{eq:-2}
\end{equation}
where ${\rm Tr}\left({\bf A}_{l}^{T}\boldsymbol{\Sigma}^{-1}{\bf A}_{l}\right)$
measures the difference between $\boldsymbol{\Sigma}$ and the positive
semidefinite matrix ${\bf A}_{l}{\bf A}_{l}^{T}$. $h_{l}\left(\cdot\right)$
is, in general, an increasing function that increases the penalty
as $\boldsymbol{\Sigma}$ deviates from ${\bf A}_{l}{\bf A}_{l}^{T}$,
which is considered to be the prior target that we wish to shrink
$\boldsymbol{\Sigma}$ to.

We first give an intuitive argument on the condition that ensures
the existence of the estimator. Since the estimator $\hat{\boldsymbol{\Sigma}}$
is defined as the minimizer to the penalized loss function, it exists
if $\tilde{L}\left(\boldsymbol{\Sigma}\right)\to+\infty$ on the boundary
of $\mathbb{S}_{++}^{K}$ by Lemma \ref{lem:equivalence lem}, and
clearly $\hat{\boldsymbol{\Sigma}}$ is nonsingular. We infer $\boldsymbol{\Sigma}$
by the samples $\left\{ {\bf x}_{i}\right\} $, if the samples are
concentrated on some subspace, naturally we ``guess'' the distribution
is degenerate, i.e., $\hat{\boldsymbol{\Sigma}}$ is singular. Therefore,
the samples are required to be sufficiently spread out in the whole
space so that the inference leads to a nonsingular $\hat{\boldsymbol{\Sigma}}$.
Under the case when we have a prior information that $\boldsymbol{\Sigma}$
should be close to the matrix ${\bf A}_{l}{\bf A}_{l}^{T}$, to ensure
$\hat{\boldsymbol{\Sigma}}$ being nonsingular we need to distribute
more ${\bf x}_{i}$'s in the null space of ${\bf A}_{l}{\bf A}_{l}^{T}$
and hence less in the range of ${\bf A}_{l}{\bf A}_{l}^{T}$ . To
formalize this intuition, we give the following theorem.
\begin{thm}
\label{thm:Existence Thm}For cost function
\begin{equation}
\begin{array}{ll}
\tilde{L}\left(\boldsymbol{\Sigma}\right) & =\frac{N}{2}\log\det\left(\boldsymbol{\Sigma}\right)+\sum_{i=1}^{N}\rho\left({\bf x}_{i}^{T}\boldsymbol{\Sigma}^{-1}{\bf x}_{i}\right)\\
 & +\left(\alpha\log\det\left(\boldsymbol{\Sigma}\right)+\sum_{l=1}^{L}\alpha_{l}h_{l}\left({\rm Tr}\left({\bf A}_{l}^{T}\boldsymbol{\Sigma}^{-1}{\bf A}_{l}\right)\right)\right)
\end{array}\label{eq:-3}
\end{equation}
defined on positive definite matrices $\boldsymbol{\Sigma}\succ{\bf 0}$
with $\rho\left(\cdot\right)$ and $h\left(\cdot\right)$ being continuous
functions, define $a_{\rho}$ and $a_{\rho}^{'}$ for $\rho$, $a_{l}$
and $a_{l}^{'}$ for $\alpha_{l}h_{l}$'s according to (\ref{eq:sill})
and (\ref{eq:floor}), then $\tilde{L}\left(\boldsymbol{\Sigma}\right)\to+\infty$
on the boundary of the set $\mathbb{S}_{++}^{K}$ if the following
conditions are satisfied:

(i) no ${\bf x}_{i}$ lies on the origin;

(ii) for any proper subspace\textup{ $S$}
\[
\begin{array}{ll}
P_{N}\left(S\right) & <\min\left\{ 1-\frac{\left(N+2\alpha\right)\left(K-{\rm dim}\left(S\right)\right)-\sum_{l\in\upsilon}a_{l}}{a_{\rho}N},\right.\\
 & \left.\frac{\left(N+2\alpha\right){\rm dim}\left(S\right)-\sum_{l\in\omega}a_{l}^{'}}{a_{\rho}^{'}N}\right\}
\end{array}
\]
where sets $\omega$ and $\upsilon$ are defined as $\omega=\left\{ l|{\bf A}_{l}\subseteq S\right\} $,
$\upsilon=\left\{ l|{\bf A}_{l}\nsubseteq S\right\} $;

(iii) $\left(-\frac{N}{2}-\alpha\right)K+\frac{a_{\rho}^{'}}{2}N+\frac{1}{2}\sum_{l}a_{l}^{'}<0$
and $\frac{a_{\rho}}{2}N-\left(\frac{N}{2}+\alpha\right)K+\frac{1}{2}\sum_{l}a_{l}>0$.\end{thm}
\begin{IEEEproof}
See Appendix A.\end{IEEEproof}
\begin{rem}
Condition (i) avoids the scenario when ${\bf x}_{i}^{T}\boldsymbol{\Sigma}^{-1}{\bf x}_{i}$
takes value $0$ and $\rho\left(s\right)$ is undefined at $s=0$,
for example $\rho\left(s\right)=\log\left(s\right)$ for the log-likelihood
function. The first part in condition (ii), $P_{N}\left(S\right)<1-\frac{\left(N+2\alpha\right)\left(K-{\rm dim}\left(S\right)\right)-\sum_{l\in\upsilon}a_{l}}{a_{\rho}N}$,
ensures $\tilde{L}\left(\boldsymbol{\Sigma}\right)\to+\infty$ under
the case that some but not all eigenvalues $\lambda_{j}$ of $\boldsymbol{\Sigma}$
tend to zero, and the second part in condition (ii), $P_{N}\left(S\right)<\frac{\left(N+2\alpha\right){\rm dim}\left(S\right)-\sum_{l\in\omega}a_{l}^{'}}{a_{\rho}^{'}N}$,
ensures $\tilde{L}\left(\boldsymbol{\Sigma}\right)\to+\infty$ under
the case that some but not all eigenvalues $\lambda_{j}$ of $\boldsymbol{\Sigma}$
tend to positive infinity. Together they force $\tilde{L}\left(\boldsymbol{\Sigma}\right)\to+\infty$
when $\frac{\lambda_{\max}}{\lambda_{\min}}\to0$. The first part
of condition (iii) ensures $\tilde{L}\left(\boldsymbol{\Sigma}\right)\to+\infty$
when all $\lambda\to+\infty$ and the second part ensures $\tilde{L}\left(\boldsymbol{\Sigma}\right)\to+\infty$
when all $\lambda\to0$. \end{rem}
\begin{cor}
\label{cor:checkable condition}Assuming the population distribution
$f\left(\cdot\right)$ is continuous, and the matrices ${\bf A}_{l}$
are full rank, condition (ii) in Theorem \ref{thm:Existence Thm}
simplifies to:
\[
\left\{ \begin{array}{l}
\sum_{l}a_{l}-\left(N+2\alpha\right)\left(K-d\right)>a_{\rho}\left(d-N\right)\\
\alpha>\frac{a_{\rho}^{'}-N}{2}
\end{array},\right.\forall1\leq d\leq K-1.
\]
\end{cor}
\begin{IEEEproof}
The conclusion follows easily from the following two facts: given
that the population distribution $f\left(\cdot\right)$ is continuous,
and no ${\bf x}_{i}$ lies on the origin, any $1\leq d<K$ sample
points define a proper subspace $S$ with ${\rm dim}\left(S\right)=d$
with probability one; and since ${\bf A}_{l}$'s are full rank, the
set $\omega=\emptyset$.
\end{IEEEproof}
Under the regularity conditions provided in Theorem \ref{thm:Existence Thm},
Lemma \ref{lem:equivalence lem} implies a minimizer $\hat{\boldsymbol{\Sigma}}$
of $\tilde{L}\left(\boldsymbol{\Sigma}\right)$ exists and is positive
definite, therefore it needs to satisfy the condition $\frac{\partial\tilde{L}\left(\boldsymbol{\Sigma}\right)}{\partial\boldsymbol{\Sigma}}={\bf 0}$.

We then show how Theorem \ref{thm:Existence Thm} works for Tyler's
estimator defined as the nonsingular minimizer of (\ref{eq:Loss function Sigma}).
Notice that the loss function $L^{{\rm Tyler}}\left(\boldsymbol{\Sigma}\right)$
is scale-invariant, we have $L^{{\rm Tyler}}\left(c\boldsymbol{\Sigma}_{0}\right)=L^{{\rm Tyler}}\left(\boldsymbol{\Sigma}_{0}\right)={\rm constant}$
for any positive definite $\boldsymbol{\Sigma}_{0}$. This implies
that there are cases when $\boldsymbol{\Sigma}$ goes to the boundary
of $\mathbb{S}_{++}^{K}$ and $L^{{\rm Tyler}}\left(\boldsymbol{\Sigma}\right)$
will not go to positive infinity. Due to this reason, condition (iii)
is violated in Theorem \ref{thm:Existence Thm}. To handle the scaling
issue, we introduce a trace constraint ${\rm Tr}\left(\boldsymbol{\Sigma}\right)=1$.

For the Tyler's problem of minimizing (\ref{eq:Loss function Sigma}),
we seek for the condition that ensures $L^{{\rm Tyler}}\left(\boldsymbol{\Sigma}\right)\to+\infty$
when $\boldsymbol{\Sigma}$ goes to the boundary of the set $\left\{ \boldsymbol{\Sigma}|\boldsymbol{\Sigma}\succ{\bf 0},\ {\rm Tr}\left(\boldsymbol{\Sigma}\right)=1\right\} $
relative to $\left\{ \boldsymbol{\Sigma}|\boldsymbol{\Sigma}\succcurlyeq{\bf 0},\ {\rm Tr}\left(\boldsymbol{\Sigma}\right)=1\right\} $.
The condition implies that there is a unique minimizer $\hat{\boldsymbol{\Sigma}}$
that minimizes $L^{{\rm Tyler}}\left(\boldsymbol{\Sigma}\right)$
over the set $\left\{ \boldsymbol{\Sigma}|\boldsymbol{\Sigma}\succ{\bf 0},\ {\rm Tr}\left(\boldsymbol{\Sigma}\right)=1\right\} $,
and by it is equivalent to the existence of a unique (up to a positive
scaling factor) minimizer $\boldsymbol{\Sigma}^{\star}$ that minimizes
$L^{{\rm Tyler}}\left(\boldsymbol{\Sigma}\right)$ over the set $\mathbb{S}_{++}^{K}$
since $L^{{\rm Tyler}}\left(\boldsymbol{\Sigma}\right)$ is scale-invariant.

The constraint ${\rm Tr}\left(\boldsymbol{\Sigma}\right)=1$ excludes
the case that any of $\lambda_{j}\to+\infty$ and the case all $\lambda_{j}\to0$,
hence we only need to let $L^{{\rm Tyler}}\left(\boldsymbol{\Sigma}\right)\to+\infty$
under the case that some but not all $\lambda_{j}\to0$, which corresponds
to the condition $P_{N}\left(S\right)<1-\frac{\left(N+2\alpha\right)\left(K-{\rm dim}\left(S\right)\right)-\sum_{l\in\upsilon}a_{l}}{a_{\rho}N}$
in Theorem \ref{thm:Existence Thm}. For Tyler's cost function $L^{{\rm Tyler}}\left(\boldsymbol{\Sigma}\right)$,
we have $\rho\left(s\right)=\frac{K}{2}\log s$ and $\alpha=0$, $a_{\rho}=a_{\rho}^{'}=K$,
therefore Theorem \ref{thm:Existence Thm} leads to the condition
on the samples: $P_{N}\left(S\right)<\frac{{\rm dim}\left(S\right)}{K}$,
or $N\geq K+1$ if the population distribution  $f\left(\cdot\right)$
is continuous, which reduces to the condition given in \cite{kent1988maximum}.

\subsection{Regularization via Wiesel's penalty}

In \cite{wiesel2012unified}, Wiesel proposed a regularization penalty
$h\left(\boldsymbol{\Sigma}\right)$ that results in a shrinkage estimator.
Specifically, the penalty terms that encourage shrinkage towards an
identity matrix and more generally towards an arbitrary prior matrix
${\bf T}$ are defined as follows:
\begin{equation}
\begin{array}{l}
h^{{\rm identity}}\left(\boldsymbol{\Sigma}\right)=K\log\left({\rm Tr}\left(\boldsymbol{\Sigma}^{-1}\right)\right)+\log\det\left(\boldsymbol{\Sigma}\right)\\
h^{{\rm target}}\left(\boldsymbol{\Sigma}\right)=K\log\left({\rm Tr}\left(\boldsymbol{\Sigma}^{-1}{\bf T}\right)\right)+\log\det\left(\boldsymbol{\Sigma}\right).
\end{array}\label{eq:shrinkage-penality}
\end{equation}
As can be seen the penalty terms are scale-invariant. Wiesel justified
the choice of the above mentioned penalty functions by showing that
the minimizer of the penalty functions would be some scaled multiple
of $\boldsymbol{\mathrm{I}}$ (or ${\bf T}$). Thus adding this penalty
terms to the Tyler's cost function would yield estimators that are
shrunk towards $\boldsymbol{\mathrm{I}}$ (or ${\bf T}$). In the
rest of this subsection we consider the general case $h^{{\rm target}}$
only, where the penalty term shrinks $\boldsymbol{\Sigma}$ to scalar
multiples of ${\bf T}$, and we make the assumption that ${\bf T}$
is positive definite, which is reasonable since $\boldsymbol{\Sigma}$
must be a positive definite matrix. The cost function is restated
below for convenience
\begin{equation}
\begin{array}{ll}
L^{{\rm Wiesel}}\left(\boldsymbol{\Sigma}\right) & =\log\det\left(\boldsymbol{\Sigma}\right)+\frac{K}{N}\sum_{i=1}^{N}\log\left({\bf x}_{i}^{T}\boldsymbol{\Sigma}^{-1}{\bf x}_{i}\right)\\
 & +\alpha_{0}\left(K\log\left({\rm Tr}\left(\boldsymbol{\Sigma}^{-1}{\bf T}\right)\right)+\log\det\left(\boldsymbol{\Sigma}\right)\right).
\end{array}\label{eq: wiesel penalized loss functionl scatter}
\end{equation}
Minimizing $L^{{\rm Wiesel}}\left(\boldsymbol{\Sigma}\right)$ gives
the fixed-point condition
\begin{equation}
\boldsymbol{\Sigma}=\frac{1}{1+\alpha_{0}}\frac{K}{N}\sum_{i=1}^{N}\frac{{\bf x}_{i}{\bf x}_{i}^{T}}{{\bf x}_{i}^{T}\boldsymbol{\Sigma}^{-1}{\bf x}_{i}}+\frac{\alpha_{0}}{1+\alpha_{0}}\frac{K{\bf T}}{{\rm Tr}\left(\boldsymbol{\Sigma}^{-1}{\bf T}\right)}.\label{eq: eqn wiesel scatter}
\end{equation}

Recall that in the absence of regularization (i.e., $\alpha_{0}=0$),
a solution to the fixed-point equation exists under the condition
$P_{N}\left(S\right)<\frac{{\rm dim}\left(S\right)}{N}$. With the
regularization, however, it is not clear. We start giving a result
for the uniqueness and then come back to the existence.
\begin{thm}
\label{thm:unique wiesel scatter}If (\ref{eq: eqn wiesel scatter})
has a solution, then it is unique up to a positive scaling factor.\end{thm}
\begin{IEEEproof}
It's easy to see if $\boldsymbol{\Sigma}$ solves (\ref{eq: eqn wiesel scatter}),
$c\boldsymbol{\Sigma}$ is also a solution for $c>0$. Without loss
of generality assume $\boldsymbol{\Sigma}={\bf I}$ is a solution,
otherwise define $\tilde{{\bf x}}_{i}=\boldsymbol{\Sigma}^{-\frac{1}{2}}{\bf x}_{i}$
and $\tilde{{\bf T}}=\boldsymbol{\Sigma}^{-\frac{1}{2}}{\bf T}\boldsymbol{\Sigma}^{-\frac{1}{2}}$,
and that there exists another solution $\boldsymbol{\Sigma}_{1}$.
Denote the eigenvalues of $\boldsymbol{\Sigma}_{1}$ as $\lambda_{1}\ge\cdots\geq\lambda_{K}$
with at least one strictly inequality, then under the condition that
${\bf T}$ is positive definite
\begin{eqnarray*}
\boldsymbol{\Sigma}_{1} & = & \frac{1}{1+\alpha_{0}}\frac{K}{N}\sum_{i=1}^{N}\frac{{\bf x}_{i}{\bf x}_{i}^{T}}{{\bf x}_{i}^{T}\boldsymbol{\Sigma}_{1}^{-1}{\bf x}_{i}}+\frac{\alpha_{0}}{1+\alpha_{0}}\frac{K{\bf T}}{{\rm Tr}\left(\boldsymbol{\Sigma}_{1}^{-1}{\bf T}\right)}\\
 & \prec & \frac{1}{1+\alpha_{0}}\frac{K}{N}\sum_{i=1}^{N}\frac{{\bf x}_{i}{\bf x}_{i}^{T}}{\lambda_{1}^{-1}{\bf x}_{i}^{T}{\bf x}_{i}}+\frac{\alpha_{0}}{1+\alpha_{0}}\frac{K{\bf T}}{{\rm Tr}\left(\lambda_{1}^{-1}{\bf T}\right)}\\
 & = & \lambda_{1}{\bf I}
\end{eqnarray*}
where the inequality follows from the fact that ${\rm Tr}\left({\bf S}\boldsymbol{\Sigma}_{1}^{-1}\right)>{\rm Tr}\left(\lambda_{1}^{-1}{\bf S}\right)$
for any positive definite matrix ${\bf S}$ and the last equality
follows from the assumption that ${\bf I}$ is a solution to (\ref{eq: eqn wiesel scatter}).
We have the contradiction $\lambda_{1}<\lambda_{1}$, hence all the
eigenvalues of $\boldsymbol{\Sigma}_{1}$ should be equal, i.e., $\boldsymbol{\Sigma}_{1}=\lambda{\bf I}$.
\end{IEEEproof}
Before establishing the existence condition, we give an example when
the solution to (\ref{eq: eqn wiesel scatter}) does not exist for
illustration.
\begin{example}
Consider the case when all ${\bf x}_{i}$'s are aligned in one direction.
Eigendecompose $\boldsymbol{\Sigma}={\bf U}\boldsymbol{\Lambda}{\bf U}^{T}$
and choose ${\bf u}_{1}$ to be aligned with the ${\bf x}_{i}$'s,
let $\lambda_{1}\to+\infty$ while others $0<c\leq\lambda<+\infty$.
Ignoring the constant terms, the boundedness of $L^{{\rm Wiesel}}\left(\boldsymbol{\Sigma}\right)$
is equivalent to the boundedness of $\left(1+\alpha_{0}-K\right)\log\lambda_{1}$,
hence it is unbounded below if $\alpha_{0}<K-1$.
\end{example}
The example shows that $L^{{\rm Wiesel}}\left(\boldsymbol{\Sigma}\right)$
can be unbounded below implying that (\ref{eq: eqn wiesel scatter})
has no solution if the data are too concentrated and $\alpha_{0}$
is small. The following theorems gives the exact tradeoff between
data dispersion and the choice of $\alpha_{0}$.
\begin{thm}
\label{thm:Existence-wiesel}A unique solution to (\ref{eq: eqn wiesel scatter})
exists (up to a positive scaling factor) if the following conditions
are satisfied:

(i) no ${\bf x}_{i}$ lies on the origin;

(ii) for any proper subspace $S\subseteq\mathbb{R}^{K}$, $P_{N}\left(S\right)<\frac{\left(1+\alpha_{0}\right){\rm dim}\left(S\right)}{K}$,\\
and they are the global minima of the loss function (\ref{eq: wiesel penalized loss functionl scatter}).\end{thm}
\begin{IEEEproof}
We start by rewriting the function including a the scaling factor
$\frac{N}{2}$ w.r.t. (\ref{eq: wiesel penalized loss functionl scatter})
for convenience:
\begin{equation}
\begin{array}{ll}
L^{{\rm Wiesel}}\left(\boldsymbol{\Sigma}\right) & =\frac{N}{2}\log\det\left(\boldsymbol{\Sigma}\right)+\frac{K}{2}\sum_{i=1}^{N}\log\left({\bf x}_{i}^{T}\boldsymbol{\Sigma}^{-1}{\bf x}_{i}\right)\\
 & +\frac{\alpha_{0}N}{2}\left(K\log\left({\rm Tr}\left(\boldsymbol{\Sigma}^{-1}{\bf T}\right)\right)+\log\det\left(\boldsymbol{\Sigma}\right)\right).
\end{array}\label{eq:-4}
\end{equation}
Invoke Theorem \ref{thm:Existence Thm} with $\rho\left(s\right)=\frac{K}{2}\log\left(s\right)$,
$h_{1}\left(s\right)=K\log\left(s\right)$, $\alpha=\alpha_{1}=\frac{\alpha_{0}N}{2}$
and ${\bf A}_{1}={\bf T}^{\frac{1}{2}}$, hence $a_{\rho}=a_{\rho}^{'}=K$
and $a_{1}=a_{1}^{'}=2\alpha K$. By the same reasoning as for the
Tyler's loss function, the condition $P_{N}\left(S\right)<1-\frac{\left(N+2\alpha\right)\left(K-{\rm dim}\left(S\right)\right)-\sum_{l\in\upsilon}a_{l}}{a_{\rho}N}$,
which is $P_{N}\left(S\right)<\frac{\left(1+\alpha_{0}\right){\rm dim}\left(S\right)}{K}$
since ${\bf T}$ is full rank, ensures the existence of a unique solution
to (\ref{eq: eqn wiesel scatter}) under the constraint $\boldsymbol{\Sigma}\succ{\bf 0}$
and ${\rm Tr}\left(\boldsymbol{\Sigma}\right)=1$. Hence a unique
(up to a positive scaling factor) solution to (\ref{eq: eqn wiesel scatter})
exists on the set of $\mathbb{S}_{++}^{K}$ by the scale-invariant
property of $L^{{\rm Wiesel}}\left(\boldsymbol{\Sigma}\right)$.
\end{IEEEproof}
To make the existence condition checkable, we use Corollary \ref{cor:checkable condition},
Theorem \ref{thm:Existence-wiesel} then simplifies to $\alpha_{0}>\frac{K}{N}-1$
or, equivalently $N>\frac{K}{1+\alpha_{0}}$, from which we can see
that compared to the condition without regularization shrinkage allows
less number of samples, and the minimum number depends on $\alpha_{0}$.

At last, we show that the condition $P_{N}\left(S\right)<\frac{\left(1+\alpha_{0}\right){\rm dim}\left(S\right)}{K}$
is also necessary in the following proposition.
\begin{prop}
\label{prop:necessaity wiesel}If (\ref{eq: eqn wiesel scatter})
admits a solution on $\mathbb{S}_{++}^{K}$, then for any proper subspace
$S\subseteq\mathbb{R}^{K}$, $P_{N}\left(S\right)<\frac{\left(1+\alpha_{0}\right){\rm dim}\left(S\right)}{K}$,
provided that ${\bf T}$ is positive definite and $\alpha_{0}>0$.\end{prop}
\begin{IEEEproof}
For a proper subspace $S$, define ${\bf P}$ as the orthogonal projection
matrix associated to $S$, i.e., ${\bf P}{\bf x}={\bf x}$, $\forall{\bf x}\in S$.
Assume the solution is ${\bf I}$. Multiplying both sides of equation
(\ref{eq: eqn wiesel scatter}) by matrix ${\bf I}-{\bf P}$ and taking
the trace we have
\[
\begin{alignedat}{1} & K-{\rm dim}\left(S\right)\\
= & \frac{1}{1+\alpha_{0}}\frac{K}{N}\sum_{i=1}^{N}\frac{{\bf x}_{i}^{T}\left({\bf I}-{\bf P}\right){\bf x}_{i}}{{\bf x}_{i}^{T}{\bf x}_{i}}+\frac{\alpha_{0}K}{1+\alpha_{0}}\frac{{\rm Tr}\left({\bf T}-{\bf T}{\bf P}\right)}{{\rm Tr}\left({\bf T}\right)}
\end{alignedat}
\]
If ${\bf x}_{i}\in S$, then ${\bf x}_{i}^{T}\left({\bf I}-{\bf P}\right){\bf x}_{i}=0$,
and ${\bf x}_{i}^{T}\left({\bf I}-{\bf P}\right){\bf x}_{i}\leq{\bf x}_{i}^{T}{\bf x}_{i}$
if ${\bf x}_{i}\notin S$. Moreover, ${\rm Tr}\left({\bf T}{\bf P}\right)>0$
since ${\bf T}$ is positive definite. This therefore implies
\[
K-{\rm dim}\left(S\right)<\frac{1}{1+\alpha_{0}}\frac{K}{N}\left(N-NP_{N}\left(S\right)\right)+\frac{\alpha_{0}K}{1+\alpha_{0}}.
\]
Rearranging the terms yields $P_{N}\left(S\right)<\frac{\left(1+\alpha_{0}\right){\rm dim}\left(S\right)}{K}$.
\end{IEEEproof}

\subsection{Regularization via Kullback-Leibler Divergence Penalty}

An ideal penalty term should increase as $\boldsymbol{\Sigma}$ deviates
from the prior target ${\bf T}$. Wiesel's penalty function discussed
in the last subsection satisfies this property and, in this subsection,
we propose another penalty that has this property. The penalty that
we choose is the KL divergence between $\mathcal{N}_{\Sigma}\left({\bf 0},\boldsymbol{\Sigma}\right)$
and $\mathcal{N}_{T}\left({\bf 0},{\bf T}\right)$, i.e., two zero-mean
Gaussians with covariance matrices $\boldsymbol{\Sigma}$ and ${\bf T}$,
respectively. The formula for the KL divergence is as follows \cite{verdu2002spectral,cover2012elements}
\[
D_{KL}\left(\mathcal{N}_{T}\left\Vert \mathcal{N}_{\Sigma}\right.\right)=\frac{1}{2}\left({\rm Tr}\left(\boldsymbol{\Sigma}^{-1}{\bf T}\right)-K-\log\left(\frac{\det\left({\bf T}\right)}{\det\left(\boldsymbol{\Sigma}\right)}\right)\right).
\]
Ignoring the constant terms results in the following loss function:
\begin{equation}
\begin{array}{ll}
L^{{\rm KL}}\left(\boldsymbol{\Sigma}\right) & =\log\det\left(\boldsymbol{\Sigma}\right)+\frac{K}{N}\sum_{i=1}^{N}\log\left({\bf x}_{i}^{T}\boldsymbol{\Sigma}^{-1}{\bf x}_{i}\right)\\
 & +\alpha_{0}\left({\rm Tr}\left(\boldsymbol{\Sigma}^{-1}{\bf T}\right)+\log\det\left(\boldsymbol{\Sigma}\right)\right)
\end{array}\label{eq:P-KL-scatter}
\end{equation}
with the following fixed-point condition:
\begin{equation}
\boldsymbol{\Sigma}=\frac{1}{1+\alpha_{0}}\frac{K}{N}\sum_{i=1}^{N}\frac{{\bf x}_{i}{\bf x}_{i}^{T}}{{\bf x}_{i}^{T}\boldsymbol{\Sigma}^{-1}{\bf x}_{i}}+\frac{\alpha_{0}}{1+\alpha_{0}}{\bf T}.\label{eq:KL-Scatteronly}
\end{equation}
Unlike the penalty function discussed in the last subsection, KL divergence
penalty encourages shrinkage towards ${\bf T}$ without scaling ambiguity.
This can be easily seen, as the minimizer for the KL divergence penalty
is just ${\bf T}$. Notice that (\ref{eq:KL-Scatteronly}) is similar
to the diagonal loading in (\ref{eq:heuristic diagonal loading}),
but without the heuristic normalizing step.
\begin{thm}
\label{thm:Uniqueness KL scatter}If (\ref{eq:KL-Scatteronly}) has
a solution, then it is unique.\end{thm}
\begin{IEEEproof}
Without loss of generality, we assume $\boldsymbol{\Sigma}={\bf I}$
solves (\ref{eq:KL-Scatteronly}). Assume there is another matrix
$\boldsymbol{\Sigma}_{1}$ that solves (\ref{eq:KL-Scatteronly}),
and denote the largest eigenvalue of $\boldsymbol{\Sigma}_{1}$ as
$\lambda_{1}$ and suppose $\lambda_{1}>1$. We then have the following
contradiction:
\begin{eqnarray*}
\boldsymbol{\Sigma}_{1} & \preccurlyeq & \frac{1}{1+\alpha_{0}}\frac{K}{N}\sum_{i=1}^{N}\frac{{\bf x}_{i}{\bf x}_{i}^{T}}{\lambda_{1}^{-1}{\bf x}_{i}^{T}{\bf x}_{i}}+\frac{\alpha_{0}}{1+\alpha_{0}}{\bf T}\\
 & < & \frac{\lambda_{1}}{1+\alpha_{0}}\frac{K}{N}\sum_{i=1}^{N}\frac{{\bf x}_{i}{\bf x}_{i}^{T}}{{\bf x}_{i}^{T}{\bf x}_{i}}+\frac{\alpha_{0}\lambda_{1}}{1+\alpha_{0}}{\bf T}=\lambda_{1}{\bf I}
\end{eqnarray*}
which gives contradiction $\lambda_{1}<\lambda_{1}$, hence $\lambda_{1}\leq1$.
Similarly, suppose the smallest eigenvalue of $\boldsymbol{\Sigma}_{1}$
satisfies $\lambda_{K}<1$. We then have
\begin{eqnarray*}
\boldsymbol{\Sigma}_{1} & \succcurlyeq & \frac{1}{1+\alpha_{0}}\frac{K}{N}\sum_{i=1}^{N}\frac{{\bf x}_{i}{\bf x}_{i}^{T}}{\lambda_{K}^{-1}{\bf x}_{i}^{T}{\bf x}_{i}}+\frac{\alpha_{0}}{1+\alpha_{0}}{\bf T}\\
 & > & \frac{\lambda_{K}}{1+\alpha_{0}}\frac{K}{N}\sum_{i=1}^{N}\frac{{\bf x}_{i}{\bf x}_{i}^{T}}{{\bf x}_{i}^{T}{\bf x}_{i}}+\frac{\alpha_{0}\lambda_{K}}{1+\alpha_{0}}{\bf T}=\lambda_{K}{\bf I}
\end{eqnarray*}
which is a contradiction and hence $\lambda_{K}\geq1$, from which
$\boldsymbol{\Sigma}_{1}={\bf I}$ follows.\end{IEEEproof}
\begin{thm}
\label{thm:Exist-KL penalty}A unique solution to (\ref{eq:KL-Scatteronly})
exists, if

(i) no ${\bf x}_{i}$ lies on the origin;

(ii) $P_{N}\left(S\right)<\frac{\left(1+\alpha_{0}\right){\rm dim}\left(S\right)}{K}$,\\
and it is the global minimum of loss function (\ref{eq:P-KL-scatter}).\end{thm}
\begin{IEEEproof}
Equivalently, we can define
\begin{equation}
\begin{array}{ll}
L^{{\rm KL}}\left(\boldsymbol{\Sigma}\right) & =\frac{N}{2}\log\det\left(\boldsymbol{\Sigma}\right)+\frac{K}{2}\sum_{i=1}^{N}\log\left({\bf x}_{i}^{T}\boldsymbol{\Sigma}^{-1}{\bf x}_{i}\right)\\
 & +\frac{\alpha_{0}N}{2}\left({\rm Tr}\left(\boldsymbol{\Sigma}^{-1}{\bf T}\right)+\log\det\left(\boldsymbol{\Sigma}\right)\right).
\end{array}\label{eq:-5}
\end{equation}
Invoke Theorem \ref{thm:Existence Thm} with $\rho\left(s\right)=\frac{K}{2}\log\left(s\right)$,
$h_{1}\left(s\right)=s$, $\alpha=\alpha_{1}=\frac{\alpha_{0}N}{2}$
and ${\bf A}_{1}={\bf T}^{\frac{1}{2}}$, hence $a_{\rho}=a_{\rho}^{'}=K$,
$a_{1}=+\infty$, $a_{1}^{'}=0$. Since ${\bf T}$ is full rank and
$a_{1}=+\infty$, condition (ii) reduces to $P_{N}\left(S\right)<\frac{\left(1+\alpha_{0}\right){\rm dim}\left(S\right)}{K}$.
Condition (iii) is satisfied, hence an interior minimum exists. Furthermore,
it is the unique minimum, hence it is global.\end{IEEEproof}
\begin{rem}
The only difference between the regularized estimator discussed in
this subsection and the heuristic estimator in (\ref{eq:heuristic diagonal loading})
is the extra normalizing step in (\ref{eq:heuristic diagonal loading}).
With the trace normalization, \cite{chen2011robust} proved that the
iteration implied by (\ref{eq:heuristic diagonal loading}) converges
to a unique solution without any assumption of the data. However,
the iteration implied by (\ref{eq:KL-Scatteronly}), which is based
on minimizing a negative log-likelihood function penalized via the
KL divergence function, requires some regularity conditions to be
satisfied (cf. Theorem \ref{thm:Exist-KL penalty}). According to
Corollary \ref{cor:checkable condition}, the condition simplifies
to $\alpha_{0}>\frac{K}{N}-1$ if the population distribution is continuous.\end{rem}
\begin{prop}
If (\ref{eq:KL-Scatteronly}) admits a solution on $\mathbb{S}_{++}^{K}$,
then for any proper subspace $S\subseteq\mathbb{R}^{K}$, $P_{N}\left(S\right)<\frac{\left(1+\alpha_{0}\right){\rm dim}\left(S\right)}{K}$,
provided that ${\bf T}$ is positive definite and $\alpha_{0}>0$.\end{prop}
\begin{IEEEproof}
Multiply both sides of equation (\ref{eq:KL-Scatteronly}) by ${\bf T}^{-\frac{1}{2}}$
and define $\tilde{\boldsymbol{\Sigma}}={\bf T}^{-\frac{1}{2}}\boldsymbol{\Sigma}{\bf T}^{-\frac{1}{2}}$,
$\tilde{{\bf x}}_{i}={\bf T}^{-\frac{1}{2}}{\bf x}_{i}$ yields
\begin{equation}
\tilde{\boldsymbol{\Sigma}}=\frac{1}{1+\alpha_{0}}\frac{K}{N}\sum_{i=1}^{N}\frac{\tilde{{\bf x}}_{i}\tilde{{\bf x}}_{i}^{T}}{{\bf \tilde{x}}_{i}^{T}\tilde{\boldsymbol{\Sigma}}^{-1}\tilde{{\bf x}}_{i}}+\frac{\alpha_{0}}{1+\alpha_{0}}{\bf I}.\label{eq:whitened KL fixed point eqn}
\end{equation}
The rest of the proof follows the same token as Proposition \ref{prop:necessaity wiesel}.
\end{IEEEproof}
Finally, we show in the following proposition that the Wiesel's shrinkage
estimator defined as solution to (\ref{eq: eqn wiesel scatter}) and
KL shrinkage estimator defined as solution to (\ref{eq:KL-Scatteronly})
are equivalent.
\begin{prop}
The solution to fixed point equation (\ref{eq:KL-Scatteronly}) solves
(\ref{eq: eqn wiesel scatter}) and, conversely, any solution of (\ref{eq: eqn wiesel scatter})
solves (\ref{eq:KL-Scatteronly}) with a proper scaling factor.\end{prop}
\begin{IEEEproof}
If $\alpha_{0}$ is zero, the statement is trivial. We consider the
case $\alpha_{0}\neq0$. Following the argument of previous proposition
we arrive at equation (\ref{eq:whitened KL fixed point eqn}). It
has been shown in \cite{pascal2013generalized} that the unique solution
$\tilde{\boldsymbol{\Sigma}}$ to (\ref{eq:whitened KL fixed point eqn})
satisfies ${\rm Tr}\left(\tilde{\boldsymbol{\Sigma}}^{-1}\right)=K$
given $\alpha_{0}>0$, hence ${\rm Tr}\left(\boldsymbol{\Sigma}^{-1}{\bf T}\right)=K$.
Substitute it into equation (\ref{eq: eqn wiesel scatter}) yields
exactly equation (\ref{eq:KL-Scatteronly}) with solution $\boldsymbol{\Sigma}$,
which indicates $\boldsymbol{\Sigma}$ solves (\ref{eq: eqn wiesel scatter}).
The second part of the proposition follows from the fact that Wiesel's
fixed-point equation (\ref{eq: eqn wiesel scatter}) has a unique
solution up to a positive scaling factor.
\end{IEEEproof}

\section{Algorithms}

Before going the specific algorithms, we first briefly introduce the
concepts of majorization-minimization \cite{hunter2004tutorial,razaviyayn2013unified}.
Consider the following optimization problem
\begin{equation}
\begin{array}{ll}
\underset{{\bf x}}{{\rm minimize}} & f\left({\bf x}\right)\\
{\rm subject\ to} & {\bf x}\in\mathcal{X}
\end{array}\label{eq:-6}
\end{equation}
where $f\left(\cdot\right)$ is assumed to be a continuous function,
not necessarily convex, and $\mathcal{X}$ is a closed convex set.

At a given point ${\bf x}_{t}$, the majorization-minimization algorithm
finds a surrogate function $g\left({\bf x}|{\bf x}_{t}\right)$ that
satisfies the following properties:
\begin{equation}
\begin{array}{l}
f\left({\bf x}_{t}\right)=g\left({\bf x}_{t}|{\bf x}_{t}\right)\\
f\left({\bf x}\right)\leq g\left({\bf x}|{\bf x}_{t}\right)\ \forall{\bf x}\in\mathcal{X}\\
f'\left({\bf x}_{t};{\bf d}\right)=g'\left({\bf x}_{t};{\bf d}|{\bf x}_{t}\right)\ \forall{\bf x}_{t}+{\bf d}\in\mathcal{X}
\end{array}\label{eq:-7}
\end{equation}
with $f'\left({\bf x};{\bf d}\right)$ stands for directional derivative.
The surrogate function $g\left({\bf x}|{\bf x}_{t}\right)$ is assumed
to be continuous in ${\bf x}$ and ${\bf x}_{t}$%
\footnote{Notice that if both $f\left({\bf x}\right)$ and $g\left({\bf x}|{\bf x}_{t}\right)$
are continuously differentiable, then the first two conditions in
(\ref{eq:-7}) imply the third.%
}.

The majorization-minimization algorithm updates ${\bf x}$ as
\[
{\bf x}_{t+1}=\arg\min_{{\bf x}\in\mathcal{X}}g\left({\bf x}|{\bf x}_{t}\right).
\]
It is proved that every limit point of the sequence $\left\{ {\bf x}_{t}\right\} $
converges to a stationary point of problem (\ref{eq:-6}), and under
the assumption that the level set $\mathcal{X}^{0}=\left\{ {\bf x}|f\left({\bf x}\right)\leq f\left({\bf x}_{0}\right)\right\} $
is compact, the distance between $\left\{ {\bf x}_{t}\right\} $ and
the set of stationary points reduces to zero in the limit\cite{razaviyayn2013unified}.

In the rest of this section, for any continuous differentiable function
$f\left({\bf y}\right)$, we define $f\left({\bf y}\right)=+\infty$
when $\lim_{{\bf x}\to{\bf y}}f\left({\bf x}\right)=+\infty$.

\subsection{Regularization via Wiesel's Penalty}

In \cite{wiesel2012unified}, Wiesel derived Tyler's iteration (\ref{eq:Alg-Tyler Scatter})
but without the trace normalization step, from the majorization-minimization
perspective, with surrogate function $g\left(\boldsymbol{\Sigma}|\boldsymbol{\Sigma}_{t}\right)$
for (\ref{eq:Loss function Sigma}) defined as
\begin{equation}
g\left(\boldsymbol{\Sigma}|\boldsymbol{\Sigma}_{t}\right)=\frac{N}{2}\log\det\left(\boldsymbol{\Sigma}\right)+\sum_{i=1}^{N}\frac{K}{2}\frac{{\bf x}_{i}^{T}\boldsymbol{\Sigma}^{-1}{\bf x}_{i}}{{\bf x}_{i}^{T}\boldsymbol{\Sigma}_{t}^{-1}{\bf x}_{i}}+{\rm const.}\label{eq:-8}
\end{equation}
A positive definite stationary point of $g\left(\boldsymbol{\Sigma}|\boldsymbol{\Sigma}_{t}\right)$
satisfies the first equation of (\ref{eq:Alg-Tyler Scatter}). By
the same technique, to solve the problem
\begin{equation}
\begin{array}{ll}
\underset{\boldsymbol{\Sigma}}{{\rm minimize}} & \log\det\left(\boldsymbol{\Sigma}\right)+\frac{K}{N}\sum_{i=1}^{N}\log\left({\bf x}_{i}^{T}\boldsymbol{\Sigma}^{-1}{\bf x}_{i}\right)\\
 & +\alpha_{0}\left(K\log\left({\rm Tr}\left(\boldsymbol{\Sigma}^{-1}{\bf T}\right)\right)+\log\det\left(\boldsymbol{\Sigma}\right)\right)\\
{\rm subject\ to} & \boldsymbol{\Sigma}\succcurlyeq{\bf 0}.
\end{array}\label{eq:P-wiesel penalty}
\end{equation}
Wiesel derived the iteration (\ref{eq:wiesel penality}) by majorizing
(\ref{eq: wiesel penalized loss functionl scatter}) with function
\begin{equation}
\begin{aligned}\left(1+\alpha_{0}\right)\log\det & \left(\boldsymbol{\Sigma}\right)+\frac{K}{N}\sum_{i=1}^{N}\frac{{\bf x}_{i}^{T}\boldsymbol{\Sigma}^{-1}{\bf x}_{i}}{{\bf x}_{i}^{T}\boldsymbol{\Sigma}_{t}^{-1}{\bf x}_{i}}\\
 & +\frac{\alpha_{0}K}{{\rm Tr}\left(\boldsymbol{\Sigma}_{t}^{-1}{\bf T}\right)}{\rm Tr}\left(\boldsymbol{\Sigma}^{-1}{\bf T}\right)+{\rm const}.
\end{aligned}
\label{eq:partial majorize}
\end{equation}

It is worth pointing out that if we do the change of variable $\boldsymbol{\psi}=\boldsymbol{\Sigma}^{-1}$
in $L^{{\rm Wiesel}}\left(\boldsymbol{\Sigma}\right)$ and linearize
the term $\log\left({\bf x}_{i}^{T}\boldsymbol{\Sigma}^{-1}{\bf x}_{i}\right)$,
this also leads to the same iteration (\ref{eq:wiesel penality}).

In the rest of this subsection, we prove the convergence of the iteration
(\ref{eq:wiesel penality}) proposed by Wiesel, but with an additional
trace normalization step, i.e., our modified iteration takes the form:
\begin{equation}
\begin{array}{l}
\tilde{\boldsymbol{\Sigma}}_{t+1}=\frac{1}{1+\alpha_{0}}\frac{K}{N}\sum_{i=1}^{N}\frac{{\bf x}_{i}{\bf x}_{i}^{T}}{{\bf x}_{i}^{T}\boldsymbol{\Sigma}_{t}^{-1}{\bf x}_{i}}+\frac{\alpha_{0}}{1+\alpha_{0}}\frac{K{\bf T}}{{\rm Tr}\left(\boldsymbol{\Sigma}_{t}^{-1}{\bf T}\right)}\\
\boldsymbol{\Sigma}_{t+1}=\frac{\tilde{\boldsymbol{\Sigma}}_{t+1}}{{\rm Tr}\left(\tilde{\boldsymbol{\Sigma}}_{t+1}\right)}.
\end{array}\label{eq:sigma partial majorize}
\end{equation}
Denote the set $\mathcal{S}=\left\{ \boldsymbol{\Sigma}|{\rm Tr}\left(\boldsymbol{\Sigma}\right)=1,\ \boldsymbol{\Sigma}\succcurlyeq{\bf 0}\right\} $.

\begin{algorithm}
\begin{enumerate}
\item Initialize $\boldsymbol{\Sigma}_{0}$ as an arbitrary positive definite
matrix.
\item Do iteration
\[
\begin{array}{l}
\tilde{\boldsymbol{\Sigma}}_{t+1}=\frac{1}{1+\alpha_{0}}\frac{K}{N}\sum_{i=1}^{N}\frac{{\bf x}_{i}{\bf x}_{i}^{T}}{{\bf x}_{i}^{T}\boldsymbol{\Sigma}_{t}^{-1}{\bf x}_{i}}+\frac{\alpha_{0}}{1+\alpha_{0}}\frac{K{\bf T}}{{\rm Tr}\left(\boldsymbol{\Sigma}_{t}^{-1}{\bf T}\right)}\\
\boldsymbol{\Sigma}_{t+1}=\frac{\tilde{\boldsymbol{\Sigma}}_{t+1}}{{\rm Tr}\left(\tilde{\boldsymbol{\Sigma}}_{t+1}\right)}
\end{array}
\]
until convergence.
\end{enumerate}
\caption{\label{alg:Wiesel's-Iteration}Wiesel's shrinkage estimator }
\end{algorithm}

\begin{lem}
The set $\mathcal{X}^{0}=\left\{ \boldsymbol{\Sigma}|L^{{\rm Wiesel}}\left(\boldsymbol{\Sigma}\right)\leq L^{{\rm Wiesel}}\left(\boldsymbol{\Sigma}_{0}\right)\right\} \cap\mathcal{S}$
is a compact set.\end{lem}
\begin{IEEEproof}
${\rm Tr}\left(\boldsymbol{\Sigma}\right)=1$ implies the set $\mathcal{X}^{0}$
is bounded. The set is closed follows easily from the fact that $L^{{\rm Wiesel}}\left(\boldsymbol{\Sigma}\right)\to+\infty$
when $\boldsymbol{\Sigma}$ tends to be singular.\end{IEEEproof}
\begin{lem}
\label{lem:unique mini surrogate}The $\tilde{\boldsymbol{\Sigma}}_{t+1}$
given in (\ref{eq:sigma partial majorize}) is the unique minimizer
to surrogate function (\ref{eq:partial majorize}).\end{lem}
\begin{IEEEproof}
For surrogate function (\ref{eq:partial majorize}), its value goes
to positive infinity when $\frac{\lambda_{\max}\left(\boldsymbol{\Sigma}\right)}{\lambda_{\min}\left(\boldsymbol{\Sigma}\right)}\to+\infty$,
since it majorizes $L^{{\rm Wiesel}}\left(\boldsymbol{\Sigma}\right)$
and $L^{{\rm Wiesel}}\left(\boldsymbol{\Sigma}\right)\to+\infty$
in this case. Now consider the case when $\frac{\lambda_{\max}}{\lambda_{\min}}=O\left(1\right)$.
Define $\bar{\boldsymbol{\Sigma}}=\frac{\boldsymbol{\Sigma}}{\boldsymbol{\Sigma}_{K,K}}$,
then function (\ref{eq:partial majorize}) can be rewritten as

\begin{equation}
\begin{aligned} & \left(1+\alpha_{0}\right)\left(\log\det\left(\bar{\boldsymbol{\Sigma}}\right)+K\log\lambda_{\min}\right)\\
 & +\frac{K}{N}\sum_{i=1}^{N}\lambda_{\min}^{-1}\frac{{\bf x}_{i}^{T}\bar{\boldsymbol{\Sigma}}^{-1}{\bf x}_{i}}{{\bf x}_{i}^{T}\boldsymbol{\Sigma}_{t}^{-1}{\bf x}_{i}}+\frac{\alpha_{0}K}{{\rm Tr}\left(\boldsymbol{\Sigma}_{t}^{-1}{\bf T}\right)}{\rm Tr}\left(\lambda_{\min}^{-1}\bar{\boldsymbol{\Sigma}}^{-1}{\bf T}\right)+{\rm const}.
\end{aligned}
\label{eq:partial majorize-rewrite}
\end{equation}
The terms $\log\det\left(\bar{\boldsymbol{\Sigma}}\right)$, $\frac{{\bf x}_{i}^{T}\bar{\boldsymbol{\Sigma}}^{-1}{\bf x}_{i}}{{\bf x}_{i}^{T}\boldsymbol{\Sigma}_{t}^{-1}{\bf x}_{i}}$
and ${\rm Tr}\left(\bar{\boldsymbol{\Sigma}}^{-1}{\bf T}\right)$
are all constants bounded away from both $0$ and $+\infty$. It is
easy to see that when $\lambda_{\min}\to0$ or $\lambda_{\min}\to+\infty$,
(\ref{eq:partial majorize-rewrite}) goes to $+\infty$. Therefore
we conclude that the value of Wiesel's surrogate function (\ref{eq:partial majorize})
goes to $+\infty$ when $\boldsymbol{\Sigma}$ approaches the boundary
of $\mathbb{S}_{+}^{K}$. The fact that $\tilde{\boldsymbol{\Sigma}}_{t+1}$
given in (\ref{eq:sigma partial majorize}) is the unique solution
to the stationary equation implies that it is the unique minimizer
of (\ref{eq:partial majorize}) on the set $\mathbb{S}_{+}^{K}$.\end{IEEEproof}
\begin{prop}
The sequence $\left\{ \boldsymbol{\Sigma}_{t}\right\} $ generated
by Algorithm \ref{alg:Wiesel's-Iteration} converges to the global
minimizer of problem (\ref{eq:P-wiesel penalty}).\end{prop}
\begin{IEEEproof}
It is proved in Theorem \ref{thm:Existence-wiesel} that under the
conditions provided in Theorem \ref{thm:Existence-wiesel}, the minimizer
$\hat{\boldsymbol{\Sigma}}$ for problem
\begin{equation}
\begin{array}{ll}
\underset{\boldsymbol{\Sigma}\succ{\bf 0}}{{\rm minimize}} & \log\det\left(\boldsymbol{\Sigma}\right)+\frac{K}{N}\sum_{i=1}^{N}\log\left({\bf x}_{i}^{T}\boldsymbol{\Sigma}^{-1}{\bf x}_{i}\right)\\
 & +\alpha_{0}\left(K\log\left({\rm Tr}\left(\boldsymbol{\Sigma}^{-1}{\bf T}\right)\right)+\log\det\left(\boldsymbol{\Sigma}\right)\right)\\
{\rm subject\ to} & {\rm Tr}\left(\boldsymbol{\Sigma}\right)=1
\end{array}\label{eq:-15}
\end{equation}
exists and is unique, furthermore, it solves problem (\ref{eq:P-wiesel penalty}).
It is also proved that the objective function $L^{{\rm Wiesel}}\left(\boldsymbol{\Sigma}\right)\to+\infty$
on the boundary of the set $\mathcal{S}$. We now show that the sequence
$\left\{ \boldsymbol{\Sigma}_{t}\right\} $ converges to unique minimizer
of (\ref{eq:-15}).

Denote the surrogate function in general as $g\left(\boldsymbol{\Sigma}|\boldsymbol{\Sigma}_{t}\right)$,
by Lemma \ref{lem:unique mini surrogate} we therefore have the following
inequality
\begin{eqnarray*}
L^{{\rm Wiesel}}\left(\boldsymbol{\Sigma}_{t}\right) & = & g\left(\boldsymbol{\Sigma}_{t}|\boldsymbol{\Sigma}_{t}\right)\geq g\left(\tilde{\boldsymbol{\Sigma}}_{t+1}|\boldsymbol{\Sigma}_{t}\right)\\
 & \geq & L^{{\rm Wiesel}}\left(\tilde{\boldsymbol{\Sigma}}_{t+1}\right)=L^{{\rm Wiesel}}\left(\boldsymbol{\Sigma}_{t+1}\right),
\end{eqnarray*}
which means $\left\{ L^{{\rm Wiesel}}\left(\boldsymbol{\Sigma}_{t}\right)\right\} $
is a non-increasing sequence.

Assume that there exists converging subsequence $\boldsymbol{\Sigma}_{t_{j}}\to\boldsymbol{\Sigma}_{\infty}$,
then
\[
\begin{aligned} & g\left(\boldsymbol{\Sigma}|\boldsymbol{\Sigma}_{t_{j}}\right)\geq g\left(\tilde{\boldsymbol{\Sigma}}_{t_{j}+1}|\boldsymbol{\Sigma}_{t_{j}}\right)\geq L^{{\rm Wiesel}}\left(\tilde{\boldsymbol{\Sigma}}_{t_{j}+1}\right)\\
= & L^{{\rm Wiesel}}\left(\boldsymbol{\Sigma}_{t_{j}+1}\right)\geq L^{{\rm Wiesel}}\left(\boldsymbol{\Sigma}_{t_{j+1}}\right)=g\left(\boldsymbol{\Sigma}_{t_{j+1}}|\boldsymbol{\Sigma}_{t_{j+1}}\right),\\
 & \forall\boldsymbol{\Sigma}\succ{\bf 0}.
\end{aligned}
\]
Letting $j\to+\infty$ results in
\[
g\left(\boldsymbol{\Sigma}|\boldsymbol{\Sigma}_{\infty}\right)\geq g\left(\boldsymbol{\Sigma}_{\infty}|\boldsymbol{\Sigma}_{\infty}\right)\ \forall\boldsymbol{\Sigma}\succ{\bf 0},
\]
which implies that the directional derivative $L^{{\rm Wiesel}'}\left(\boldsymbol{\Sigma}_{\infty};\boldsymbol{\Delta}\right)\geq{\bf 0},\forall\boldsymbol{\Sigma}_{\infty}+\boldsymbol{\Delta}\succ{\bf 0}$.
The limit $\boldsymbol{\Sigma}_{\infty}$ is nonsingular since if
$\boldsymbol{\Sigma}_{\infty}$ is singular $L^{{\rm Wiesel}}\left(\boldsymbol{\Sigma}_{\infty}\right)=+\infty$,
but $L^{{\rm Wiesel}}\left(\boldsymbol{\Sigma}_{\infty}\right)\leq L^{{\rm Wiesel}}\left(\boldsymbol{\Sigma}_{0}\right)<+\infty$
given that $\boldsymbol{\Sigma}_{0}\succ{\bf 0}$, which is a contradiction.
Since $\boldsymbol{\Sigma}_{\infty}\succ{\bf 0}$ and the function
is continuously differentiable, we have $\frac{\partial L^{{\rm Wiesel}}\left(\boldsymbol{\Sigma}\right)}{\partial\boldsymbol{\Sigma}}\Big|_{\boldsymbol{\Sigma}_{\infty}}={\bf 0}$.
Since ${\rm Tr}\left(\boldsymbol{\Sigma}_{\infty}\right)=1$, $\boldsymbol{\Sigma}_{\infty}=\hat{\boldsymbol{\Sigma}}$.

The set $\mathcal{X}^{0}=\left\{ \boldsymbol{\Sigma}|L\left(\boldsymbol{\Sigma}\right)\leq L\left(\boldsymbol{\Sigma}_{0}\right)\right\} \cap\mathcal{S}$
is a compact set, and $\left\{ \boldsymbol{\Sigma}_{t}\right\} $
lies in this set, hence $\left\{ \boldsymbol{\Sigma}_{t}\right\} $
converges to $\hat{\boldsymbol{\Sigma}}$.
\end{IEEEproof}

\subsection{Regularization via Kullback-Leibler Penalty}

Following the same approach, for the KL divergence penalty problem:
\begin{equation}
\begin{array}{ll}
\underset{\boldsymbol{\Sigma}}{{\rm minimize}} & \log\det\left(\boldsymbol{\Sigma}\right)+\frac{K}{N}\sum_{i=1}^{N}\log\left({\bf x}_{i}^{T}\boldsymbol{\Sigma}^{-1}{\bf x}_{i}\right)\\
 & +\alpha_{0}\left({\rm Tr}\left(\boldsymbol{\Sigma}^{-1}{\bf T}\right)+\log\det\left(\boldsymbol{\Sigma}\right)\right)\\
{\rm subject\ to} & \boldsymbol{\Sigma}\succcurlyeq{\bf 0}
\end{array}\label{eq:P-KL}
\end{equation}
We can majorize $L^{{\rm KL}}\left(\boldsymbol{\Sigma}\right)$ at
$\boldsymbol{\Sigma}_{t}$ by function
\begin{equation}
\left(1+\alpha_{0}\right)\log\det\left(\boldsymbol{\Sigma}\right)+\frac{K}{N}\sum_{i=1}^{N}\frac{{\bf x}_{i}^{T}\boldsymbol{\Sigma}^{-1}{\bf x}_{i}}{{\bf x}_{i}^{T}\boldsymbol{\Sigma}_{t}^{-1}{\bf x}_{i}}+\alpha_{0}{\rm Tr}\left(\boldsymbol{\Sigma}^{-1}{\bf T}\right)\label{eq:surrogate-partial linearize}
\end{equation}
the stationary condition leads to the iteration
\begin{equation}
\boldsymbol{\Sigma}_{t+1}=\frac{1}{1+\alpha_{0}}\frac{K}{N}\sum_{i=1}^{N}\frac{{\bf x}_{i}{\bf x}_{i}^{T}}{{\bf x}_{i}^{T}\boldsymbol{\Sigma}_{t}^{-1}{\bf x}_{i}}+\frac{\alpha_{0}}{1+\alpha_{0}}{\bf T}.\label{eq:-12}
\end{equation}
Algorithm \ref{alg:KL iteration} summarizes the procedure for KL
shrinkage estimator.

\begin{algorithm}
\begin{enumerate}
\item Initialize $\boldsymbol{\Sigma}_{0}$ as an arbitrary positive definite
matrix.
\item Do iteration
\[
\begin{array}{l}
\boldsymbol{\Sigma}_{t+1}=\frac{1}{1+\alpha_{0}}\frac{K}{N}\sum_{i=1}^{N}\frac{{\bf x}_{i}{\bf x}_{i}^{T}}{{\bf x}_{i}^{T}\boldsymbol{\Sigma}_{t}^{-1}{\bf x}_{i}}+\frac{\alpha_{0}}{1+\alpha_{0}}{\bf T}\end{array}
\]
until convergence.
\end{enumerate}
\caption{\label{alg:KL iteration}KL divergence penalized shrinkage estimator}
\end{algorithm}

\begin{prop}
The sequence $\left\{ \boldsymbol{\Sigma}_{t}\right\} $ generated
by Algorithm \ref{alg:KL iteration} converges to the global minimizer
of problem (\ref{eq:P-KL}).\end{prop}
\begin{IEEEproof}
We verify the assumptions required for the convergence of algorithm
\cite{razaviyayn2013unified}, namely (\ref{eq:-7}) and the compactness
of initial level set $\mathcal{X}^{0}=\left\{ \boldsymbol{\Sigma}|L^{{\rm KL}}\left(\boldsymbol{\Sigma}\right)<L^{{\rm KL}}\left(\boldsymbol{\Sigma}_{0}\right),\ \boldsymbol{\Sigma}\succ{\bf 0}\right\} $.

The first condition in (\ref{eq:-7}) is satisfied by construction.
To verify the second condition, we see that the gradient of the surrogate
function $g\left(\boldsymbol{\Sigma}|\boldsymbol{\Sigma}_{t}\right)$
has a unique zero. Since $g\left(\boldsymbol{\Sigma}|\boldsymbol{\Sigma}_{t}\right)$
is a global upperbound for $L^{{\rm KL}}\left(\boldsymbol{\Sigma}\right)$,
$g\left(\boldsymbol{\Sigma}|\boldsymbol{\Sigma}_{t}\right)\to+\infty$
as $\boldsymbol{\Sigma}$ goes to the boundary of $\mathbb{S}_{+}^{K}$.
By the continuity of $g\left(\boldsymbol{\Sigma}|\boldsymbol{\Sigma}_{t}\right)$,
a minimizer $\boldsymbol{\Sigma}^{\star}\succ{\bf 0}$ exists and
has to satisfy $\frac{\partial g}{\partial\boldsymbol{\Sigma}}={\bf 0}$.
Therefore the unique zero has to be the global minimum, i.e., $\boldsymbol{\Sigma}_{t+1}=\arg\min_{\boldsymbol{\Sigma}\succcurlyeq{\bf 0}}g\left(\boldsymbol{\Sigma}|\boldsymbol{\Sigma}_{t}\right)$.
The last condition is satisfied since $L^{{\rm KL}}\left(\boldsymbol{\Sigma}\right)$
is continuously differentiable on $\mathbb{S}_{++}^{K}$.

It is proved in Theorems \ref{thm:Uniqueness KL scatter} and \ref{thm:Exist-KL penalty}
that on set $\mathbb{S}_{++}^{K}$, $L^{{\rm KL}}\left(\boldsymbol{\Sigma}\right)$
has a unique stationary point and it is the global minimum. Furthermore,
the conditions in Theorem \ref{thm:Exist-KL penalty} ensures $L^{{\rm KL}}\left(\boldsymbol{\Sigma}\right)\to+\infty$
when $\boldsymbol{\Sigma}$ goes to the boundary of $\mathbb{S}_{+}^{K}$.
The initial set $\mathcal{X}^{0}=\left\{ \boldsymbol{\Sigma}|L^{{\rm KL}}\left(\boldsymbol{\Sigma}\right)<L^{{\rm KL}}\left(\boldsymbol{\Sigma}_{0}\right),\ \boldsymbol{\Sigma}\succ{\bf 0}\right\} $
is compact follows easily.

Therefore the sequence $\left\{ \boldsymbol{\Sigma}_{t}\right\} $
converges to the set of stationary points, hence the global minimum
of problem (\ref{eq:P-KL}).
\end{IEEEproof}

\subsection{Estimation with Structure Constraints}

In this subsection, we briefly discuss the covariance estimation problem
with structure constraints. In general, the uniqueness of the estimator
cannot be guaranteed. However, algorithms can still be derived based
on majorization-minimization when the constraint set $\mathcal{C}$
is convex. In this case, we can majorize the objective functions $L^{{\rm Wiesel}}\left(\boldsymbol{\Sigma}\right)$
and $L^{{\rm KL}}\left(\boldsymbol{\Sigma}\right)$ by
\[
\begin{aligned} & g^{{\rm Wiesel}}\left(\boldsymbol{\Sigma}|\boldsymbol{\Sigma}_{t}\right)\\
= & \left(1+\alpha_{0}\right){\rm Tr}\left(\boldsymbol{\Sigma}_{t}^{-1}\boldsymbol{\Sigma}\right)+\frac{K}{N}\sum_{i=1}^{N}\frac{{\bf x}_{i}^{T}\boldsymbol{\Sigma}^{-1}{\bf x}_{i}}{{\bf x}_{i}^{T}\boldsymbol{\Sigma}_{t}^{-1}{\bf x}_{i}}+\frac{\alpha_{0}K{\rm Tr}\left(\boldsymbol{\Sigma}^{-1}{\bf T}\right)}{{\rm Tr}\left(\boldsymbol{\Sigma}_{t}^{-1}{\bf T}\right)}
\end{aligned}
\]
and
\[
\begin{aligned} & g^{{\rm KL}}\left(\boldsymbol{\Sigma}|\boldsymbol{\Sigma}_{t}\right)\\
= & \left(1+\alpha_{0}\right){\rm Tr}\left(\boldsymbol{\Sigma}_{t}^{-1}\boldsymbol{\Sigma}\right)+\frac{K}{N}\sum_{i=1}^{N}\frac{{\bf x}_{i}^{T}\boldsymbol{\Sigma}^{-1}{\bf x}_{i}}{{\bf x}_{i}^{T}\boldsymbol{\Sigma}_{t}^{-1}{\bf x}_{i}}+\alpha_{0}{\rm Tr}\left(\boldsymbol{\Sigma}^{-1}{\bf T}\right)
\end{aligned}
\]
respectively, ignoring the constant term. Without any additional constraint,
setting the gradient of $g\left(\cdot\right)$ to zero yields update
\[
\boldsymbol{\Sigma}_{t+1}=\boldsymbol{\Sigma}_{t}^{\frac{1}{2}}\left(\boldsymbol{\Sigma}_{t}^{-\frac{1}{2}}{\bf M}_{t}\boldsymbol{\Sigma}_{t}^{-\frac{1}{2}}\right)^{1/2}\boldsymbol{\Sigma}_{t}^{\frac{1}{2}}
\]
where
\[
{\bf M}_{t}^{{\rm Wiesel}}=\frac{1}{1+\alpha_{0}}\frac{K}{N}\sum_{i=1}^{N}\frac{{\bf x}_{i}{\bf x}_{i}^{T}}{{\bf x}_{i}^{T}\boldsymbol{\Sigma}_{t}^{-1}{\bf x}_{i}}+\frac{\alpha_{0}}{1+\alpha_{0}}\frac{K{\bf T}}{{\rm Tr}\left(\boldsymbol{\Sigma}_{t}^{-1}{\bf T}\right)}
\]
and
\[
{\bf M}_{t}^{{\rm KL}}=\frac{1}{1+\alpha_{0}}\frac{K}{N}\sum_{i=1}^{N}\frac{{\bf x}_{i}{\bf x}_{i}^{T}}{{\bf x}_{i}^{T}\boldsymbol{\Sigma}_{t}^{-1}{\bf x}_{i}}+\frac{\alpha_{0}}{1+\alpha_{0}}{\bf T}.
\]
Notice that ${\bf M}_{t}$ is exactly the update we derived by only
majorizing the $\log{\rm Tr}\left(\cdot\right)$ terms in the previous
subsection, and $\boldsymbol{\Sigma}_{t+1}$ is the geometric mean
between matrices $\boldsymbol{\Sigma}_{t}$ and ${\bf M}_{t}$ \cite{zhang2013majorization}.
Intuitively $\boldsymbol{\Sigma}_{t+1}$ can be viewed as a smoothed
update of $\boldsymbol{\Sigma}_{t}$.

However, when constrained, a closed-form solution for $\boldsymbol{\Sigma}_{t+1}$
cannot be obtained in general. The surrogate function $g\left(\cdot\right)$
is convex since ${\rm {\rm Tr}\left(\boldsymbol{\Sigma}_{t}^{-1}\boldsymbol{\Sigma}\right)}$
is linear and ${\rm Tr}\left(\boldsymbol{\Sigma}^{-1}{\bf T}\right)$
is convex, $\boldsymbol{\Sigma}_{t+1}=\arg\min_{\boldsymbol{\Sigma}\in\mathcal{C}}g\left(\boldsymbol{\Sigma}|\boldsymbol{\Sigma}_{t}\right)$
can be found numerically if $\mathcal{C}$ is convex. We consider
two such examples.

\subsubsection{Covariance Matrix with Toeplitz Structure}

Toeplitz structure arises frequently in various signal processing
related fields. For example, in time series analysis, the autocovariance
matrix of a stationary process is Toeplitz. Imposing the Toeplitz
structure on $\boldsymbol{\Sigma}$ we need to solve
\[
\begin{array}{ll}
\underset{\boldsymbol{\Sigma}}{{\rm minimize}} & g\left(\boldsymbol{\Sigma}|\boldsymbol{\Sigma}_{t}\right)\\
{\rm subject\ to} & \boldsymbol{\Sigma}\succcurlyeq{\bf 0}\\
 & \boldsymbol{\Sigma}_{ij}=\boldsymbol{\Sigma}_{i+1,j+1}\ \forall i,j=1,\ldots,K-1
\end{array}
\]
for each iteration. The additional constraint is linear.

\subsubsection{Linear Additive Structure}

Suppose $\boldsymbol{\Sigma}$ can be decomposed as $\boldsymbol{\Sigma}={\bf S}+{\rm diag}\left(\sigma_{1},\ldots,\sigma_{K}\right)$,
where ${\bf S}\succcurlyeq{\bf 0}$ is signal covariance and ${\rm diag}\left(\sigma_{1},\ldots,\sigma_{K}\right)$
with $\sigma_{i}\in\mathcal{I}_{i}$ is noise covariance restricted
to some interval. Then, at each iteration we solve
\[
\begin{array}{ll}
\underset{\boldsymbol{\Sigma},{\bf S},\left\{ \sigma_{i}\right\} }{{\rm minimize}} & g\left(\boldsymbol{\Sigma}|\boldsymbol{\Sigma}_{t}\right)\\
{\rm subject\ to} & \boldsymbol{\Sigma}\succcurlyeq{\bf 0}\\
 & {\bf S}\succcurlyeq{\bf 0}\\
 & \boldsymbol{\Sigma}={\bf S}+{\rm diag}\left(\sigma_{1},\ldots,\sigma_{K}\right)\\
 & \sigma_{i}\in\mathcal{I}_{i}.
\end{array}
\]
The additional constraint is convex.

\subsection{Parameter Tuning}

A crucial issue in regularized covariance estimator is to choose the
penalty parameter $\alpha_{0}$. We have shown that if the population
distribution is continuous, for both Wiesel's penalty and KL divergence
penalty, we require $\alpha_{0}>\frac{K}{N}-1$ to guarantee the existence
of the regularized estimator.

There is a rich literature discussing the rules of parameter tuning
developed for specific estimators. A standard way is to select $\alpha_{0}$
by cross-validation, method based on random matrix theory has also
been investigated in a recent paper \cite{couillet2014large}.

\section{Numerical Results}

In all of the simulations, the estimator performance is evaluated
according to the criteria in \cite{wiesel2012unified}, namely, the
normalized mean-square error
\[
{\rm NMSE}=\frac{E\left(\left\Vert \hat{\boldsymbol{\Sigma}}-\boldsymbol{\Sigma}^{{\rm true}}\right\Vert _{F}^{2}\right)}{\left\Vert \boldsymbol{\Sigma}^{{\rm true}}\right\Vert _{F}^{2}}
\]
where all matrices $\boldsymbol{\Sigma}$ are all normalized by their
trace. The expected value is approximated by $100$ times Monte-Carlo
simulations.

The first two simulations aims at illustrating the existence conditions
for both Wiesel's shrinkage estimator and KL shrinkage estimator.
We choose $N=8$ and $K=10$ with the samples drawn a Gaussian distribution
$\mathcal{N}\left({\bf 0},\boldsymbol{\Sigma}_{0}\right)$, where
$\boldsymbol{\Sigma}_{0}$ is a randomly generated positive definite
covariance matrix. The shrinkage target ${\bf T}$ is also an arbitrary
positive definite matrix. According to the result in Section III,
$\alpha_{0}>\frac{K}{N}-1$, i.e., $\alpha_{0}>0.25$, is the necessary
and sufficient condition for the existence of a positive definite
estimator. We simulate two scenarios with $\alpha_{0}=0.24$ and $0.26$.
Fig. 1 plots $\left\Vert \boldsymbol{\Sigma}_{t}-\boldsymbol{\Sigma}_{t-1}\right\Vert _{F}$
and the inverse of the condition number, namely $\frac{\lambda_{\min}\left(\boldsymbol{\Sigma}_{t}\right)}{\lambda_{\max}\left(\boldsymbol{\Sigma}_{t}\right)}$,
as a function of the number of iterations in log-scale for Wiesel's
shrinkage estimator and with $\alpha_{0}=0.24$ (left) and $\alpha_{0}=0.26$
(right) respectively. Fig. 1 shows that for Wiesel's shrinkage estimator,
when $\alpha_{0}=0.24$ $\boldsymbol{\Sigma}_{t}$ diverges, and when
$\alpha_{0}=0.26$ $\boldsymbol{\Sigma}_{t}$ converges to a nonsingular
limit. Fig. 2 shows similar situation happens for KL shrinkage estimator.

\begin{figure}
\begin{centering}
\hspace{-0.7cm}\subfloat[]{\begin{centering}
\includegraphics[scale=0.3]{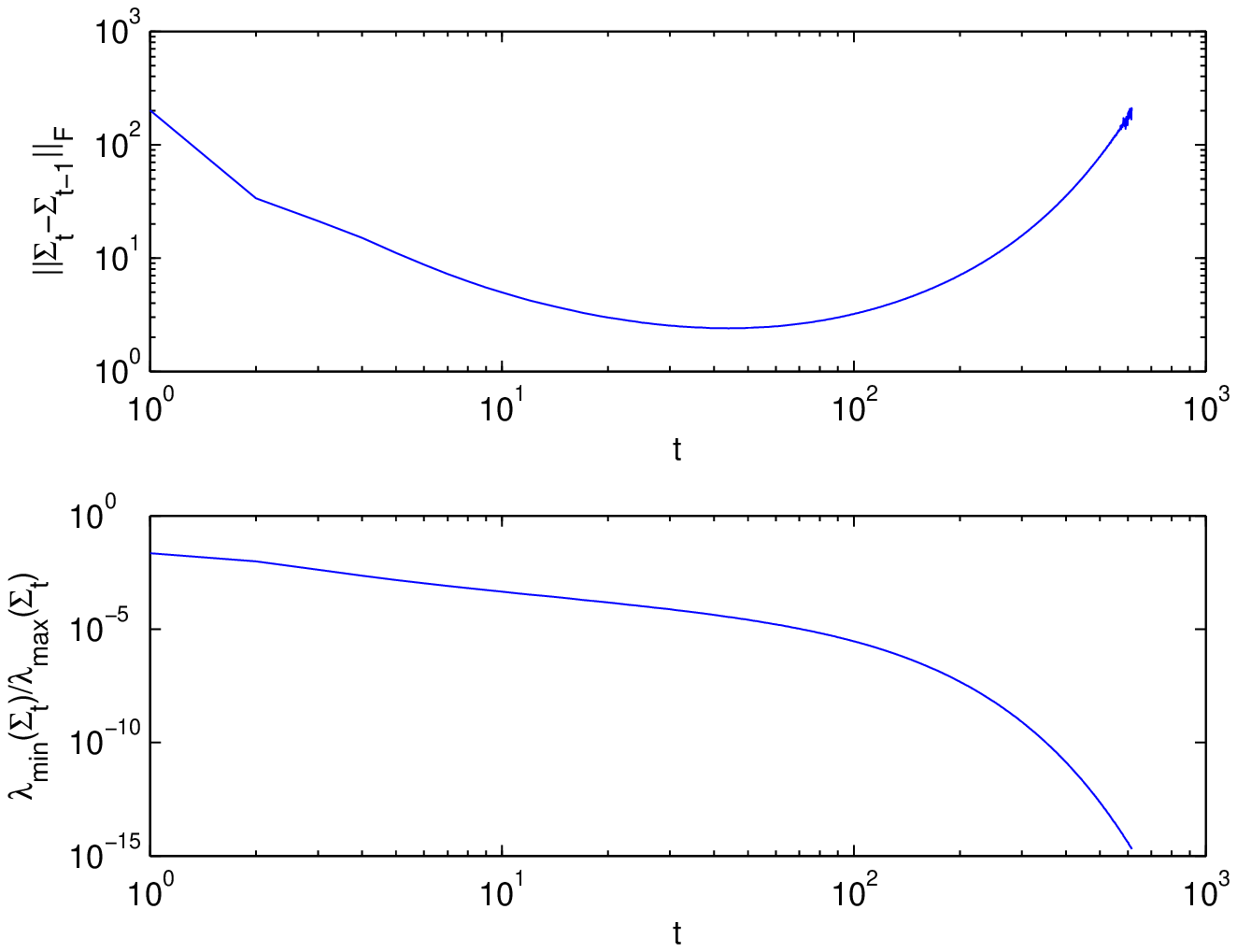}
\par\end{centering}

}\subfloat[]{\begin{centering}
\includegraphics[scale=0.3]{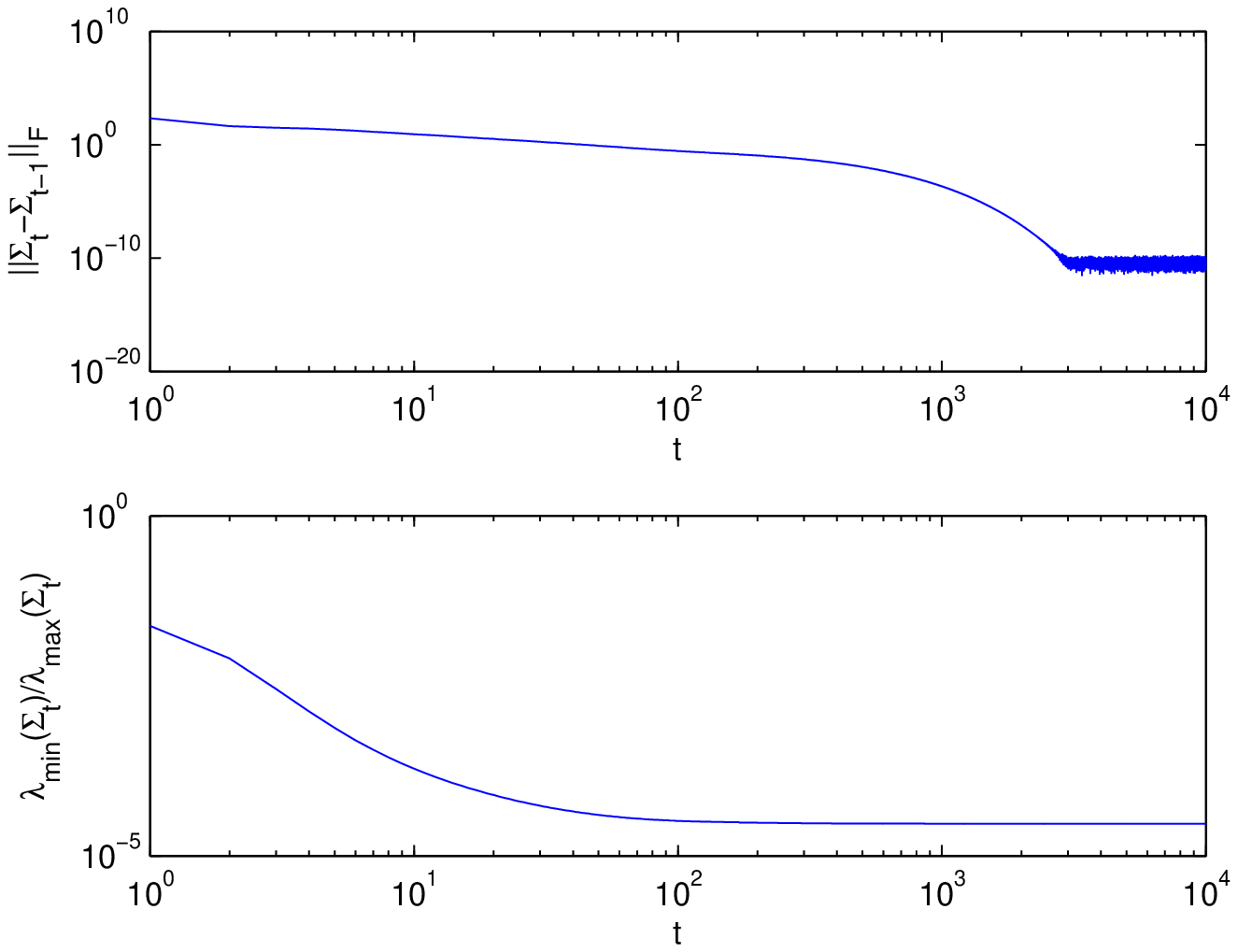}
\par\end{centering}

}
\par\end{centering}

\caption{Algorithm convergence of Wiesel's shrinkage estimator: (a) when the
existence conditions are not satisfied with $\alpha_{0}=0.24$, and
(b) when the existence conditions are satisfied with $\alpha_{0}=0.26$.}
\end{figure}

\begin{figure}
\begin{centering}
\hspace{-0.7cm}\subfloat[]{\begin{centering}
\includegraphics[scale=0.3]{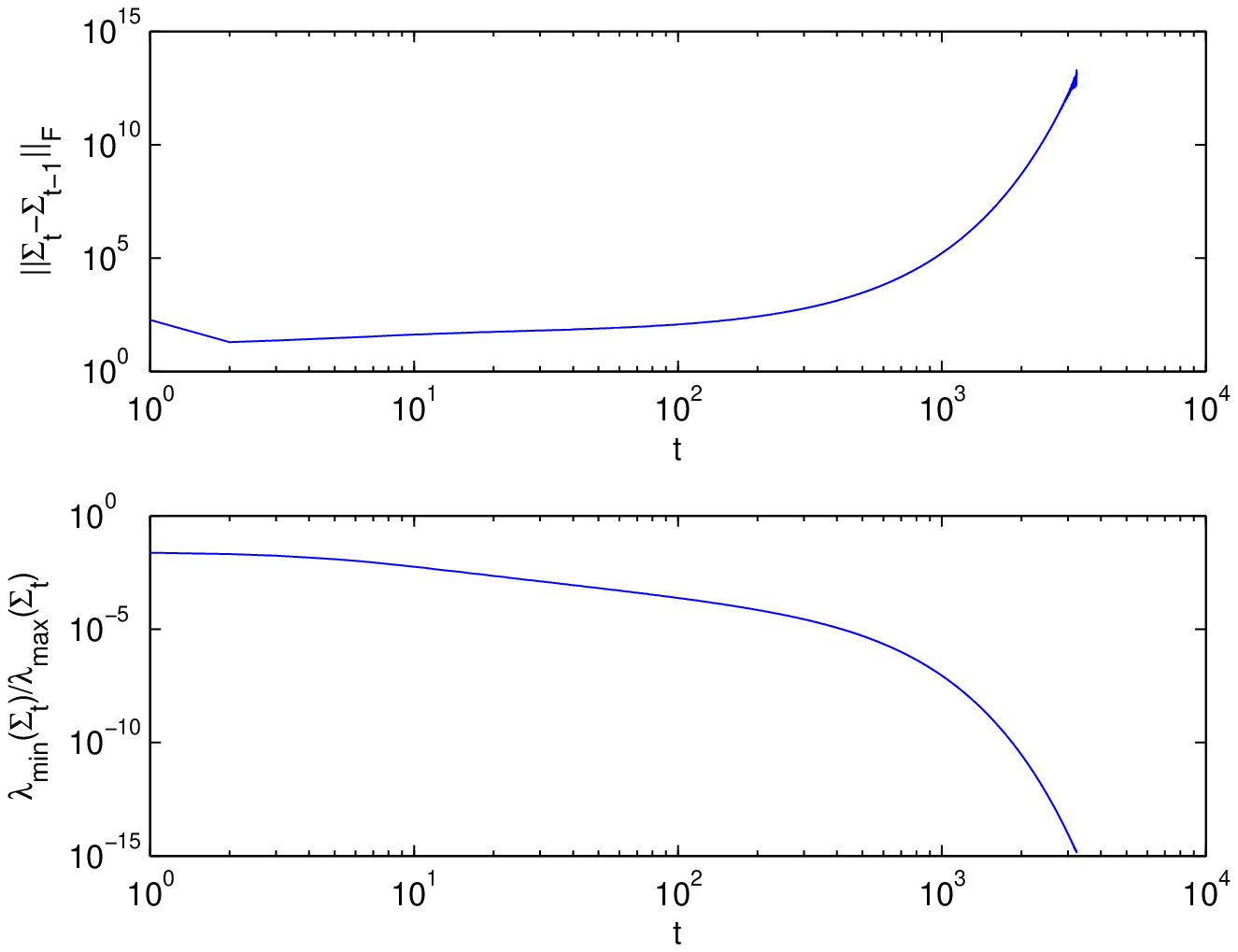}
\par\end{centering}

}\subfloat[]{\begin{centering}
\includegraphics[scale=0.3]{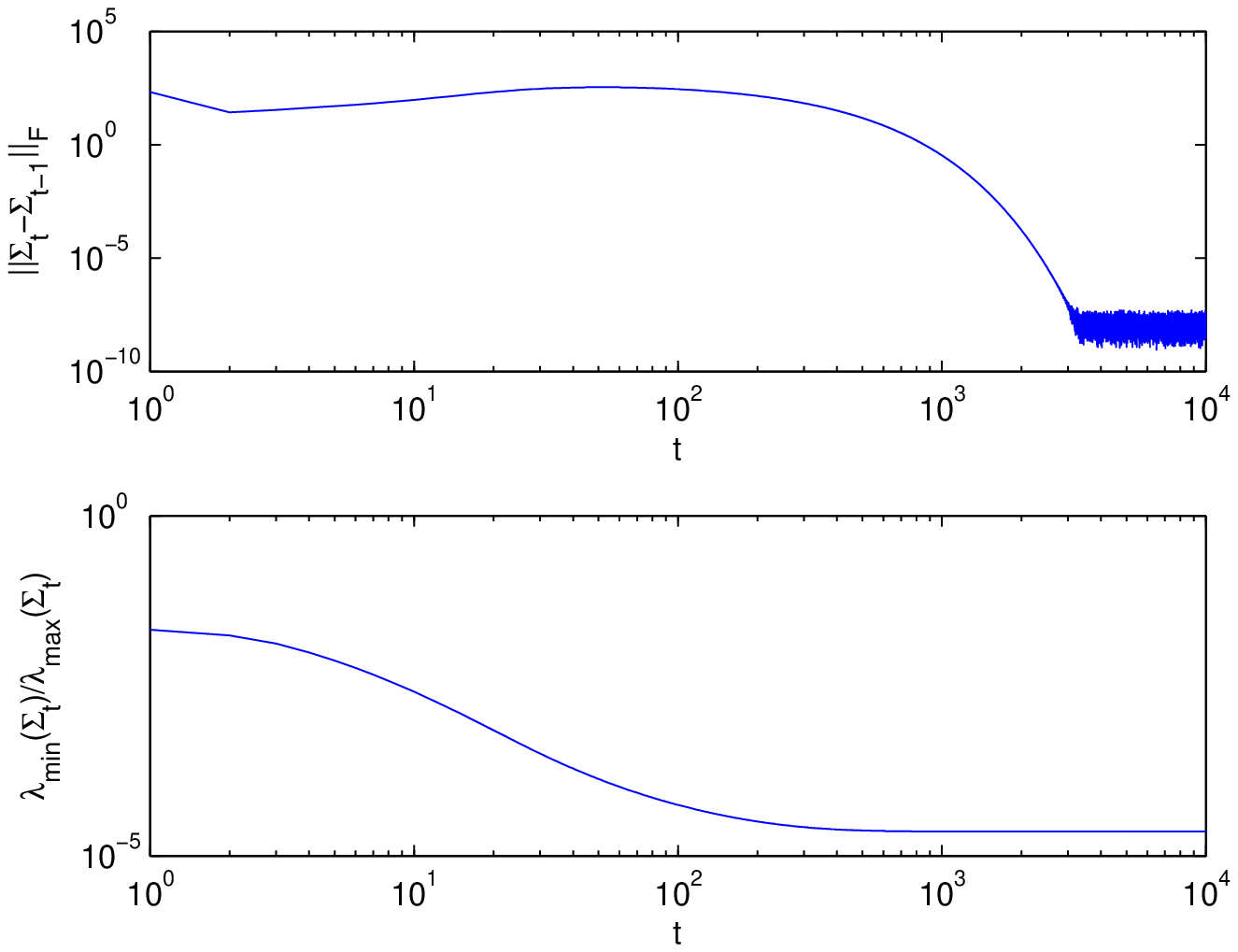}
\par\end{centering}

}
\par\end{centering}

\caption{Algorithm convergence of KL shrinkage estimator: (a) when the existence
conditions are not satisfied with $\alpha_{0}=0.24$, and (b) when
the existence conditions are satisfied with $\alpha_{0}=0.26$.}
\end{figure}

For the rest of the simulations, the shrinkage parameter $\alpha_{0}$
is selected by grid search. That is, we define $\rho=\frac{1}{1+\alpha_{0}}$
and enumerate $\rho$ uniformly on interval $\left(0,1\right]$, and
select the $\rho$ (equivalently $\alpha_{0}$) that gives the smallest
error.

\begin{figure}
\begin{centering}
\includegraphics[scale=0.5]{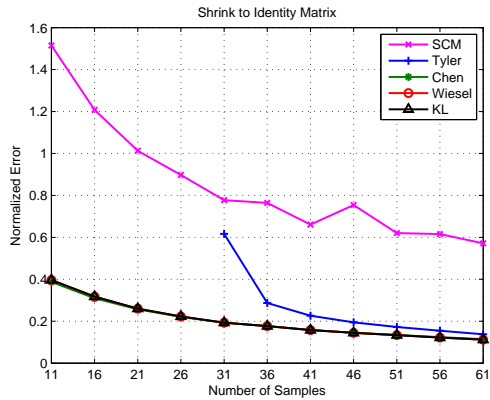}
\par\end{centering}

\caption{\label{fig:Small n large p}Illustration of the benefit of shrinkage
estimators with $K=30$ and shrinkage target matrix ${\bf I}$. }
\end{figure}

Fig. \ref{fig:Small n large p} demonstrates the performance of shrinkage
Tyler's estimator in the sample deficient case. The tuning parameter
is selected to be the one that yields the smallest NMSE for each estimator
as proposed in \cite{wiesel2012unified}. We choose the example
\[
\boldsymbol{\Sigma}\left(\beta\right)_{ij}=\beta^{\left|i-j\right|}
\]
with $K=30$. In this simulation, the underlying distribution is chosen
to be a Student's \textsl{t}-distribution with parameters $\boldsymbol{\mu}_{0}={\bf 0}$,
$\boldsymbol{\Sigma}_{0}=\boldsymbol{\Sigma}\left(0.8\right)$, and
$\nu=3$, and the shrinkage target is set to be an identity matrix.
The number of samples $N$ starts from $11$ to $61$. The curve corresponding
to Tyler's estimator starts at $N=31$ since the condition for Tyler's
estimator to exist is $N>K$, i.e., $N>30$ in this case. The figure
illustrates that both Tyler's estimator and shrinkage Tyler's estimator
outperform the sample covariance matrix when all of them exist, shrinkage
estimators exist even when $N\leq K$ and, moreover, achieve the best
performance in all cases.

\begin{figure}
\begin{centering}
\includegraphics[scale=0.5]{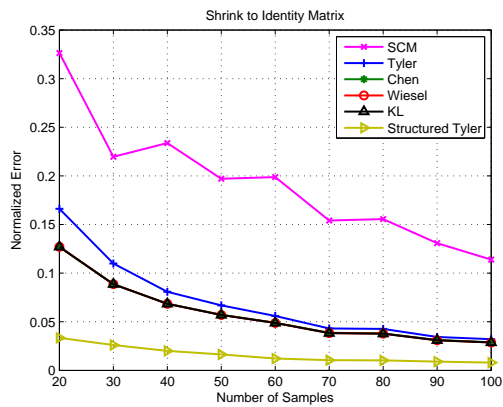}
\par\end{centering}

\caption{\label{fig:Performance-comparison-shrinkage-identity}Illustration
of the benefit of shrinkage estimators with $K=10$ and shrinkage
target matrix ${\bf I}$.}

\end{figure}

\begin{figure}
\begin{centering}
\includegraphics[scale=0.5]{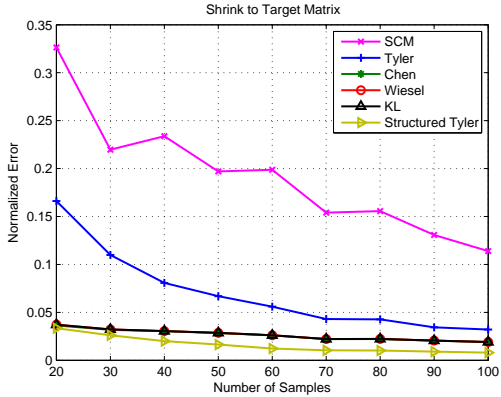}
\par\end{centering}

\caption{\label{fig:Performance-comparison-shrinkage-target}Illustration of
the benefit of shrinkage estimators with $K=10$ and a knowledge-aided
shrinkage target matrix ${\bf T}$.}
\end{figure}

Fig. \ref{fig:Performance-comparison-shrinkage-identity} and \ref{fig:Performance-comparison-shrinkage-target}
compare the performance of different shrinkage Tyler's estimators
following roughly one of the simulation set-up in \cite{wiesel2012unified}
for a fair comparison. The samples are drawn from a Student's \textsl{t}-distribution
with parameters $\boldsymbol{\mu}_{0}={\bf 0}$, $\boldsymbol{\Sigma}_{0}=\boldsymbol{\Sigma}\left(0.8\right)$
and $\nu=3$. The number of samples $N$ varies from $20$ to $100$.
Fig. \ref{fig:Performance-comparison-shrinkage-identity} shows the
estimation error when setting ${\bf T}={\bf I}$ and Fig. \ref{fig:Performance-comparison-shrinkage-target}
shows that when setting ${\bf T}=\boldsymbol{\Sigma}\left(0.7\right)$,
the searching step size of $\rho$ is set to be $0.01$. The result
indicates that estimation accuracy is increased due to shrinkage when
the number of sample is not enough. Wiesel's shrinkage estimator and
KL shrinkage estimator yield the same NMSE. Interestingly, Chen's
shrinkage estimator gives roughly the same NMSE, although with a different
shrinkage parameter $\alpha_{0}$. Chen's and KL shrinkage estimator
thus find their advantage in practice since an easier way of choosing
$\alpha_{0}$ rather than cross-validation has been investigated in
the literature \cite{chen2011robust,couillet2014large}, a detailed
comparison of them from random matrix theory perspective has also
been provided in \cite{couillet2014large}.

In both of the simulations, we include Tyler's estimator with a Toeplitz
structure constraint as introduced in the previous section. The figures
show that the structure constrained estimator achieves relatively
better performance than all other estimators both when shrinking to
${\bf I}$ and shrinking to ${\bf T}$. Although structure constraint
can be imposed on shrinkage estimators to achieve potentially even
smaller estimation error, we leave out this simulation due to the
heavy computational cost introduced both by a lack of a closed-form
solution per iteration (a SDP need to be solved numerically) and grid
searching for the best regularization parameter. The problem of accelerating
the algorithm and investigating the effect of imposing structure constraint
on shrinkage estimator are left for future work.

Finally, the performance of Tyler's estimator is tested on a real
financial data set. We choose daily close prices $p_{t}$ from Jan
1, 2008 to July 31, 2011, 720 days in total, of $K=45$ stocks from
the Hang Seng Index provided by Yahoo Finance. The samples are constructed
as $r_{t}=\log p_{t}-\log p_{t-1}$, i.e., the daily log-returns.
The process $r_{t}$ is assumed to be stationary. The vector ${\bf r}_{t}$
is constructed by stacking the log-returns of all $K$ stocks. ${\bf r}_{t}$
that is close to ${\bf 0}$ (all elements are less than $10^{-6}$)
is discarded. We compare the performance of different covariance estimators
in the minimum variance portfolio set up, that is, we allocate the
portfolio weights to minimize the overall risk. The problem can be
formulated formally as
\begin{equation}
\begin{array}{ll}
\underset{{\bf w}}{{\rm minimize}} & {\bf w}^{T}\boldsymbol{\Sigma}{\bf w}\\
{\rm subject\ to} & {\bf 1}^{T}{\bf w}=1
\end{array}\label{P:portfolio opt}
\end{equation}
with $\boldsymbol{\Sigma}$ being the covariance matrix of ${\bf r}_{t}$.
Clearly the scaling of $\boldsymbol{\Sigma}$ does not affect the
solution to this problem.

The simulation takes the following procedure. For nonshrinkage estimators,
at day $t$, we take the ${\bf r}_{i}$'s with $i\in\left[t-N^{{\rm train}}-N^{{\rm val}},t-1\right]$
as samples to estimate the normalized covariance matrix $\boldsymbol{\Sigma}$.
For a particular shrinkage estimator, the target matrix is set to
be ${\bf I}$ and the tuning parameter $\rho$ is chosen as follows:
for each value of $\rho\in\left\{ 0.01,0.02,\ldots,1\right\} $, we
calculate the shrinkage estimator $\boldsymbol{\Sigma}^{\rho}$ with
samples ${\bf r}_{i}$, $i\in\left[t-N^{{\rm train}}-N^{{\rm val}},t-N^{{\rm val}}-1\right]$
and the corresponding ${\bf w}^{\rho}$ by solving (\ref{P:portfolio opt}).
We then take the ${\bf r}_{i}$'s with $i\in\left[t-N^{{\rm val}},t-1\right]$
as validation data and evaluate the variance of portfolio series $\left\{ \left({\bf w}^{\rho}\right)^{T}{\bf r}_{i}\right\} $
in this period, the best $\rho^{\star}$ is chosen to be the one that
yields the smallest variance. Finally the shrinkage estimator is obtained
using samples ${\bf r}_{i}$ with $i\in\left[t-N^{{\rm train}}-N^{{\rm val}},t-1\right]$
and tuning parameter $\rho^{\star}$. With the allocation strategy
${\bf w}$ for each of the estimators as the solution to (\ref{P:portfolio opt}),
we construct portfolio for the next $N^{{\rm test}}$ days and collect
the returns. The procedure is repeated every $N^{{\rm test}}$ days
till the end and the variance of the portfolio constructed based on
different estimators is calculated.

\begin{figure}
\begin{centering}
\includegraphics[scale=0.5]{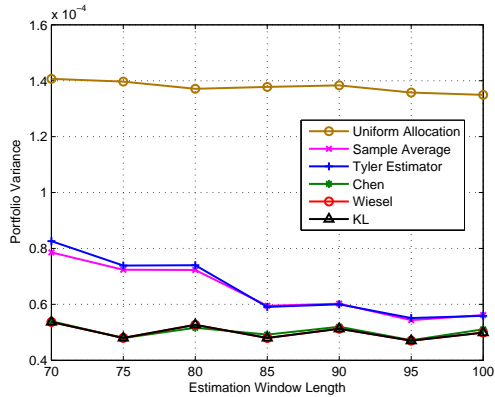}
\par\end{centering}

\caption{\label{fig:Comparison-of-portfolio}Comparison of portfolio risk constructed
based on different covariance estimators.}
\end{figure}

In the simulation, we choose $N^{{\rm val}}=N^{{\rm test}}=10$ and
vary $N^{{\rm train}}$ from $70$ to $100$. Fig. \ref{fig:Comparison-of-portfolio}
compares the variance (risk) of portfolio constructed based on different
estimators, with one additional baseline portfolio constructed by
equal investment in each asset. From the figure we can see shrinkage
estimators achieves relatively better performance than the nonshrinkage
ones.

\section{Conclusion}

In this work, we have given a rigorous proof for the existence and
uniqueness of the regularized Tyler's estimator proposed in \cite{wiesel2012unified},
and justified the heuristic diagonal loading shrinkage estimator in
\cite{4217907} by KL divergence. Under the condition that samples
are reasonably spread out, i.e., $P_{N}\left(S\right)<\frac{\left(1+\alpha_{0}\right){\rm dim}\left(S\right)}{K}$,
or $N>\frac{K}{1+\alpha_{0}}$ if the underlying distribution is continuous,
the estimators have been shown to exist and unique (up to a positive
scaling factor for Wiesel's estimator). Algorithms based on the majorization-minimization
framework have also been provided with guaranteed convergence. Finally
we have discussed structure constrained estimation and have shown
in the simulation that imposing such constraint helps improving estimation
accuracy.

\section{Appendix}

\subsection{Proof for Theorem \ref{thm:Existence Thm}}

For the loss function
\[
\begin{array}{ll}
\tilde{L}\left(\boldsymbol{\Sigma}\right) & =\frac{N}{2}\log\det\left(\boldsymbol{\Sigma}\right)+\sum_{i=1}^{N}\rho\left({\bf x}_{i}^{T}\boldsymbol{\Sigma}^{-1}{\bf x}_{i}\right)\\
 & +\left(\alpha\log\det\left(\boldsymbol{\Sigma}\right)+\sum_{l=1}^{L}\alpha_{l}h_{l}\left({\rm Tr}\left({\bf A}_{l}^{T}\boldsymbol{\Sigma}^{-1}{\bf A}_{l}\right)\right)\right)
\end{array}
\]
where the regularization term is written in general as $\alpha\log\det\left(\boldsymbol{\Sigma}\right)+\sum_{l=1}^{L}\alpha_{l}h_{l}\left({\rm Tr}\left({\bf A}_{l}^{T}\boldsymbol{\Sigma}^{-1}{\bf A}_{l}\right)\right)$.
Define $a_{\rho}$, $a_{\rho}^{'}$ for $\rho\left(s\right)$ and
$a_{l}$, $a_{l}^{'}$ for $\alpha_{l}h_{l}$ as in Definition \ref{Def: als}.

Define function
\begin{eqnarray*}
G\left(\boldsymbol{\Sigma}\right) & = & \exp\left\{ -\tilde{L}\left(\boldsymbol{\Sigma}\right)\right\} \\
 & = & \det\left(\boldsymbol{\Sigma}\right)^{-\frac{N}{2}-\alpha}\prod_{i}g_{\rho}\left({\bf x}_{i}^{T}\boldsymbol{\Sigma}^{-1}{\bf x}_{i}\right)\cdot\\
 &  & \prod g_{l}\left(\sum_{j}\lambda_{j}^{-1}\left\Vert \tilde{{\bf a}}_{lj}\right\Vert ^{2}\right)
\end{eqnarray*}
where $\tilde{{\bf a}}_{lj}$ is defined as the $j$th row of $\tilde{{\bf A}}_{l}={\bf U}^{T}{\bf A}_{l}$
with ${\bf U}$ being the unitary matrix such that ${\bf U}\boldsymbol{\Lambda}{\bf U}^{T}=\boldsymbol{\Sigma}$,
$\boldsymbol{\Lambda}={\rm diag}\left(\lambda_{1},\ldots,\lambda_{K}\right)$,
and $g_{\rho}\left(s\right)=\exp\left\{ -\rho\left(s\right)\right\} $,
$g_{l}\left(s\right)=\exp\left\{ -\alpha_{l}h_{l}\left(s\right)\right\} $.
The eigenvalues $\lambda_{j}$ is arranged in descending order, i.e.,
$\lambda_{1}\geq\cdots\geq\lambda_{K}$, and denote the inverse of
$\lambda$ as $\varphi$, hence $\varphi_{1}\le\cdots\le\varphi_{K}$.

Denote the eigenvectors corresponding to $\lambda_{j}$ as ${\bf u}_{j}$,
the subspace spanned by $\left\{ {\bf u}_{1},\ldots,{\bf u}_{j}\right\} $
as $S_{j}$ and $D_{j}=S_{j}\backslash S_{j-1}=\left\{ {\bf x}\in\mathbb{R}^{K}|{\bf x}\in S_{j},{\bf x}\notin S_{j-1}\right\} $
with $S_{0}=\left\{ 0\right\} $ and $D_{0}=\left\{ 0\right\} $.
By definition, $D_{j},\ j=0,\ldots,K$ partition the whole $\mathbb{R}^{K}$
space. Notice that $P_{N}\left\{ S_{0}\right\} =0$ by the assumption
that no ${\bf x}_{i}$ lies on the origin, we have $\sum_{j=1}^{m}P_{N}\left(D_{j}\right)=P_{N}\left(S_{m}\right)$
and $\sum_{j=m}^{K}P_{N}\left(D_{j}\right)=1-P_{N}\left(S_{m-1}\right)$.

Partition the samples ${\bf x}_{i}$ according to $D_{j}$, $j=0$
is excluded hereafter, define function
\[
\begin{array}{cc}
G_{j} & =\left\{ \begin{array}{c}
\lambda_{j}^{-\frac{N}{2}-\alpha}\prod_{{\bf x}_{i}\in D_{j}}g_{\rho}\left({\bf x}_{i}^{T}\boldsymbol{\Sigma}^{-1}{\bf x}_{i}\right)\ if\ \exists{\bf x}_{i}\in D_{j}\\
\lambda_{j}^{-\frac{N}{2}-\alpha}\ if\ no\ {\bf x}_{i}\in D_{j}
\end{array}\right.\end{array}
\]
and we have $G\left(\boldsymbol{\Sigma}\right)=\prod_{j=1}^{K}G_{j}\left(\boldsymbol{\Sigma}\right)\prod g_{l}\left(\sum_{j}\lambda_{j}^{-1}\left\Vert \tilde{{\bf a}}_{lj}\right\Vert ^{2}\right)$.

For the ${\bf A}_{l}$'s, denote ${\bf A}_{l}=\left[{\bf a}_{l1},{\bf a}_{l2},\ldots,{\bf a}_{lp}\right]$.
For each ${\bf a}_{l}$, there exists some $D_{j}$ such that ${\bf a}_{l}\in D_{j}$,
since the $D_{j}$'s partition the whole space. Define $q_{l}$ to
be the maximum index of $D_{j}$ that the ${\bf a}_{l}$'s belongs
to. Therefore we have $\left\Vert \tilde{{\bf a}}_{lq_{l}}\right\Vert \neq0$
and $\left\Vert \tilde{{\bf a}}_{lj}\right\Vert =0$ for $j>q_{l}$.

We analyze the behavior of $G\left(\boldsymbol{\Sigma}\right)$ at
the boundary of its feasible set $\mathbb{S}_{++}^{K}$, by Lemma
\ref{lem:equivalence lem}, we only need to ensure $G\left(\boldsymbol{\Sigma}\right)\to0$,
then there exists $\tilde{L}\left(\hat{\boldsymbol{\Sigma}}\right)\leq\tilde{L}\left(\boldsymbol{\Sigma}\right)\ \forall\boldsymbol{\Sigma}\succ{\bf 0}$,
and $\hat{\boldsymbol{\Sigma}}\succ{\bf 0}$.

Consider the general case that some of the $\lambda_{j}$'s go to
zero, some remains bounded away from both $0$ and positive infinity,
and the rests tend to positive infinity. Formally, define two integers
$r$ and $s$ that $1\leq r\leq s\leq K$, such that $\lambda_{j}\to+\infty$
for $j\in\left[1,r\right]$, $\lambda_{j}$ is bounded for $j\in\left(r,s\right]$
and $\lambda_{j}\to0$ for $j\in\left(s,K\right]$. Denote some arbitrary
small positive quantity by $\epsilon$.

First we analyze the terms $G_{j}$ with $\lambda_{j}\to0$. Consider
the samples ${\bf x}_{i}\in D_{h}$ for some $h\in\left(s,K\right]$,
then ${\bf x}_{i}^{T}\boldsymbol{\Sigma}^{-1}{\bf x}_{i}=\sum_{j=1}^{h}\lambda_{j}^{-1}\left\Vert {\bf u}_{j}^{T}{\bf x}_{i}\right\Vert ^{2}\geq\lambda_{h}^{-1}\left\Vert {\bf u}_{h}^{T}{\bf x}_{i}\right\Vert ^{2}$,
which is $+\infty>\left({\bf x}_{i}^{T}\boldsymbol{\Sigma}^{-1}{\bf x}_{i}\right)\lambda_{h}>0$.
Since $\lambda_{h}\to0$, we have ${\bf x}_{i}^{T}\boldsymbol{\Sigma}^{-1}{\bf x}_{i}\to+\infty$.
By definition
\[
\begin{array}{ll}
 & \lim_{\lambda_{h}\to0}g_{\rho}\left({\bf x}_{i}^{T}\boldsymbol{\Sigma}^{-1}{\bf x}_{i}\right)\left({\bf x}_{i}^{T}\boldsymbol{\Sigma}^{-1}{\bf x}_{i}\right)^{\left(a_{\rho}-\epsilon\right)/2}\\
= & \lim_{\lambda_{h}\to0}\left\{ g_{\rho}\left({\bf x}_{i}^{T}\boldsymbol{\Sigma}^{-1}{\bf x}_{i}\right)\lambda_{h}^{-\left(a_{\rho}-\epsilon\right)/2}\right\} \cdot\\
 & \left\{ \left(\left({\bf x}_{i}^{T}\boldsymbol{\Sigma}^{-1}{\bf x}_{i}\right)\lambda_{h}\right)^{\left(a_{\rho}-\epsilon\right)/2}\right\} \\
= & 0
\end{array}
\]
which implies $\lim_{\lambda_{h}\to0}\left\{ g_{\rho}\left({\bf x}_{i}^{T}\boldsymbol{\Sigma}^{-1}{\bf x}_{i}\right)\lambda_{h}^{-\left(a_{\rho}-\epsilon\right)/2}\right\} =0$,
i.e., $g_{\rho}\left({\bf x}_{i}^{T}\boldsymbol{\Sigma}^{-1}{\bf x}_{i}\right)=o\left(\lambda_{h}^{\frac{a_{\rho}-\epsilon}{2}}\right)$.
Therefore, if ${\bf x}_{i}\in D_{j}$, $g_{\rho}\left({\bf x}_{i}^{T}\boldsymbol{\Sigma}^{-1}{\bf x}_{i}\right)=o\left(\lambda_{j}^{\frac{a_{\rho}-\epsilon}{2}}\right)$.
For each $G_{j}$ we have
\[
G_{j}=o\left(\lambda_{j}^{\frac{a_{\rho}-\epsilon}{2}\cdot NP_{n}\left(D_{K}\right)-\frac{N}{2}-\alpha-\epsilon}\right)\ \forall j\geq s+1.
\]

In the second step, we analyze the terms $G_{j}$ with $\lambda_{j}\to+\infty$.
Consider the samples ${\bf x}_{i}\in D_{h}$ for some $h\in\left[1,r\right]$,
we have shown that $0<\left({\bf x}_{i}^{T}\boldsymbol{\Sigma}^{-1}{\bf x}_{i}\right)\lambda_{h}<+\infty$.
Since $\lambda_{h}\to+\infty$, $\left({\bf x}_{i}^{T}\boldsymbol{\Sigma}^{-1}{\bf x}_{i}\right)\to0$.
Given that $a_{\rho}^{'}>-\infty$,  $g_{\rho}\left({\bf x}_{i}^{T}\boldsymbol{\Sigma}^{-1}{\bf x}_{i}\right)=o\left(\varphi_{h}^{-\frac{a_{\rho}^{'}+\epsilon}{2}}\right)$
by
\[
\lim_{\varphi_{h}\to0}g_{\rho}\left({\bf x}_{i}^{T}\boldsymbol{\Sigma}^{-1}{\bf x}_{i}\right)\left({\bf x}_{i}^{T}\boldsymbol{\Sigma}^{-1}{\bf x}_{i}\right)^{\left(a_{\rho}^{'}+\epsilon\right)/2}=0.
\]
Therefore for each $G_{j}$ we have
\[
G_{j}=o\left(\varphi_{j}^{\frac{N}{2}+\alpha-\frac{a_{\rho}^{'}+\epsilon}{2}NP_{N}\left(D_{j}\right)-\epsilon}\right)\ \forall j\leq r.
\]

For the $G_{j}$ with $\lambda_{j}$ being some constant, it is easy
to see that $g_{\rho}\left({\bf x}_{i}^{T}\boldsymbol{\Sigma}^{-1}{\bf x}_{i}\right)=O\left(1\right)$,
which does not affect the order of $G\left(\boldsymbol{\Sigma}\right)$.

Now we have characterized the $g_{\rho}$'s, we move to the $g_{l}$'s.
Since $\left\Vert \tilde{{\bf a}}_{lq_{l}}\right\Vert \neq0$ and
$\left\Vert \tilde{{\bf a}}_{lj}\right\Vert =0$ for $j>q_{l}$, by
the same reasoning above, $g_{l}=o\left(\varphi_{q_{l}}^{-\frac{a_{l}^{'}+\epsilon}{2}}\right)$
if $q_{l}\leq r$ and $g_{l}=o\left(\lambda_{q_{l}}^{\frac{a_{l}-\epsilon}{2}}\right)$
if $q_{l}\geq s+1$. Therefore
\[
\begin{array}{ll}
G\left(\boldsymbol{\Sigma}\right) & =\prod_{j=1}^{K}G_{j}\left(\boldsymbol{\Sigma}\right)\prod g_{l}\left(\sum_{j}\lambda_{j}^{-1}\left\Vert \tilde{{\bf a}}_{lj}\right\Vert ^{2}\right)\\
 & =\prod_{j=1}^{r}o\left(\varphi_{j}^{\frac{N}{2}+\alpha-\frac{a_{\rho}^{'}+\epsilon}{2}NP_{N}\left(D_{j}\right)-\epsilon}\right)\cdot\\
 & \prod_{j=s+1}^{K}o\left(\lambda_{j}^{\frac{a_{\rho}-\epsilon}{2}\cdot NP_{N}\left(D_{j}\right)-\frac{N}{2}-\alpha-\epsilon}\right)\\
 & \prod_{\left\{ l|q_{l}\geq s+1\right\} }o\left(\lambda_{q_{l}}^{\frac{a_{l}-\epsilon}{2}}\right)\prod_{\left\{ l|q_{l}\leq r\right\} }o\left(\varphi_{q_{l}}^{-\frac{a_{l}^{'}+\epsilon}{2}}\right)
\end{array}
\]
with $\prod_{\left\{ l|q_{l}\geq s+1\right\} }$ defined to be $1$
if the set $\left\{ l|q_{l}\geq s+1\right\} $ is empty, and the same
for $\prod_{\left\{ l|q_{l}\leq r\right\} }$.

We make the following assumption:
\begin{equation}
\begin{array}{l}
\left(\frac{N}{2}+\alpha-\epsilon\right)m-\frac{a_{\rho}^{'}+\epsilon}{2}N\sum_{j=1}^{m}P_{N}\left(D_{j}\right)-\sum_{q_{l}\leq m}\frac{a_{l}^{'}+\epsilon}{2}\\
\geq0,\ \forall1\leq m\leq r\\
\frac{a_{\rho}}{2}\cdot N\sum_{j=m}^{K}P_{N}\left(D_{j}\right)-\left(\frac{N}{2}+\alpha+\epsilon\right)\left(K-m+1\right)\\
+\sum_{q_{l}\geq m}\frac{a_{l}-\epsilon}{2}\geq0,\ \forall K\geq m\geq s+1
\end{array}\label{eq:tmp Condtion}
\end{equation}
by the order $\lambda_{1}\geq\cdots\geq\lambda_{K}$ hence $\varphi_{1}\leq\cdots\leq\varphi_{K}$,
and base on the fact that
\[
\begin{array}{c}
o\left(\lambda_{1}^{\alpha_{1}}\right)o\left(\lambda_{2}^{\alpha_{2}}\right)=o\left(\lambda_{1}^{\alpha_{1}+\alpha_{2}}\right)\ if\ \alpha_{2}\geq0\\
o\left(\varphi_{1}^{\alpha_{1}}\right)o\left(\varphi_{2}^{\alpha_{2}}\right)=o\left(\varphi_{2}^{\alpha_{1}+\alpha_{2}}\right)\ if\ \alpha_{1}\geq0
\end{array}
\]
the order of $G\left(\boldsymbol{\Sigma}\right)$ is

\[
\begin{array}{ll}
G\left(\boldsymbol{\Sigma}\right) & =o\left(\varphi_{r}^{\left(\frac{N}{2}+\alpha-\epsilon\right)r-\frac{a_{\rho}^{'}+\epsilon}{2}N\sum_{j=1}^{r}P_{N}\left(D_{j}\right)-\sum_{q_{l}\leq r}\frac{a_{l}^{'}+\epsilon}{2}}\right)\cdot\\
 & o\left(\lambda_{s+1}^{\frac{a_{\rho}-\epsilon}{2}\cdot N\sum_{j=s+1}^{K}P_{N}\left(D_{j}\right)-\left(\frac{N}{2}+\alpha+\epsilon\right)\left(K-s\right)+\sum_{q_{l}\geq s+1}\frac{a_{l}-\epsilon}{2}}\right)
\end{array}
\]
and it goes to zero.

Now we simplify the assumption (\ref{eq:tmp Condtion}). Since $\sum_{j=1}^{m}P_{N}\left(D_{j}\right)=P_{N}\left(S_{m}\right)$
and $\sum_{j=m}^{K}P_{N}\left(D_{j}\right)=1-P_{N}\left(S_{m-1}\right)$,
and $r,s$ can take any value that satisfies $1\leq r\leq s<K$, we
end up with the following condition:
\[
\begin{array}{l}
\left(\frac{N}{2}+\alpha-\epsilon\right)d-\frac{a_{\rho}^{'}+\epsilon}{2}NP_{N}\left(S_{d}\right)-\sum_{q_{l}\leq d}\frac{a_{l}^{'}+\epsilon}{2}\geq0\\
\frac{a_{\rho}-\epsilon}{2}\cdot N\left(1-P_{N}\left(S_{d}\right)\right)-\left(\frac{N}{2}+\alpha+\epsilon\right)\left(K-d\right)\\
+\sum_{q_{l}\geq d+1}\frac{a_{l}-\epsilon}{2}\geq0
\end{array}
\]
for all $1\leq d\leq K-1$.

Define sets $\omega=\left\{ l|q_{l}\leq d\right\} $ and $\upsilon=\left\{ l|q_{l}>d\right\} $,
consider when $l\in\omega$, which means $q_{l}\leq d$, by the definition
of $q_{l}$, is equivalent to ${\rm range}\left({\bf A}_{l}\right)\subseteq S_{d}$,
similarly for $l\in\upsilon$, which means $q_{l}>d$, is equivalent
to ${\rm range}\left({\bf A}_{l}\right)\nsubseteq S_{d}$.

The condition should be valid for any ${\bf U}$ and $1\leq d\leq K-1$,
tidy up the expression and let $\epsilon\to0$ results in: for any
proper subspace $S$
\[
\begin{array}{ll}
P_{N}\left(S\right) & <\min\left\{ 1-\frac{\left(N+2\alpha\right)\left(K-{\rm dim}\left(S\right)\right)-\sum_{l\in\upsilon}a_{l}}{a_{\rho}N},\right.\\
 & \left.\frac{\left(N+2\alpha\right){\rm dim}\left(S\right)-\sum_{l\in\omega}a_{l}^{'}}{a_{\rho}^{'}N}\right\}
\end{array}
\]
where sets $\omega$ and $\upsilon$ are defined as $\omega=\left\{ l|{\rm range}\left({\bf A}_{l}\right)\subseteq S\right\} $,
$\upsilon=\left\{ l|{\rm range}\left({\bf A}_{l}\right)\nsubseteq S\right\} $.

For the case $r=0$, $1\leq s<K$, which means no $\lambda\to+\infty$
and some, not all $\lambda\to0$, following the same reasoning gives
condition
\[
P_{N}\left(S\right)<1-\frac{\left(N+2\alpha\right)\left(K-{\rm dim}\left(S\right)\right)-\sum_{l\in\upsilon}a_{l}}{a_{\rho}N}
\]
and for $s=K$, $1\leq r<K$, which means no $\lambda\to0$ and some,
not all $\lambda\to+\infty$, gives condition
\[
P_{N}\left(S\right)<\frac{\left(N+2\alpha\right){\rm dim}\left(S\right)-\sum_{l\in\omega}a_{l}^{'}}{a_{\rho}^{'}N}.
\]
Notice that the above two conditions are included in the first one.

And finally under the scenario that all $\lambda\to+\infty$, it's
easy to see $G\left(\boldsymbol{\Sigma}\right)=o\left(\varphi_{K}^{\left(\frac{N}{2}+\alpha-\epsilon\right)K-\frac{a_{\rho}^{'}}{2}N-\frac{1}{2}\sum_{l}\left(a_{l}^{'}+\epsilon\right)}\right)$
goes to zero if $\left(-\frac{N}{2}-\alpha\right)K+\frac{a_{\rho}^{'}}{2}N+\frac{1}{2}\sum_{l}a_{l}^{'}<0$,
and under the case that all $\lambda\to0$, $G\left(\boldsymbol{\Sigma}\right)=o\left(\lambda_{1}^{\frac{a_{\rho}}{2}\cdot N-\left(\frac{N}{2}+\alpha+\epsilon\right)K+\frac{1}{2}\sum_{l}\left(a_{l}-\epsilon\right)}\right)$
goes to zero if $\frac{a_{\rho}}{2}\cdot N-\left(\frac{N}{2}+\alpha\right)K+\frac{1}{2}\sum_{l}a_{l}>0$.

\bibliographystyle{IEEEtran}


\end{document}